\documentclass[pra,twocolumn,superscriptaddress,showpacs,amsmath,amstex,amssymb,citeautoscript]{revtex4-1}

\usepackage{xcolor}
\usepackage{microtype}
\definecolor{webblue}{rgb}{0, 0, 0.5} 

\usepackage[bookmarks=true,
bookmarksnumbered=false, 
bookmarksopen=false, 
colorlinks=true,
linkcolor=webblue]{hyperref}

\hypersetup{colorlinks=true,urlcolor=webblue,citecolor=webblue}
\usepackage{graphicx}
\usepackage{subfigure}
\usepackage{url}
\usepackage{hyperref}
\usepackage{inconsolata}
\usepackage{amsmath}
\usepackage{mathrsfs}
\usepackage[capitalize]{cleveref}
\usepackage{braket}
\usepackage{outlines}
\usepackage{enumitem}
\usepackage{soul}
\usepackage{color}

\usepackage{bm} 

\usepackage[utf8]{inputenc}
\usepackage{siunitx,booktabs}
\usepackage{multirow}
\usepackage{amssymb}

\usepackage{relsize}

\usepackage{soul}
\usepackage{tabularx}

\usepackage{amsmath}
\usepackage{amssymb}
\usepackage{bbm}
\usepackage{braket}
\usepackage{xcolor}

\usepackage{comment}
\usepackage{pifont}

\begin{document}

\footnotetext[1]{Some preexisting studies have considered moir\'e effects in kagome lattices, but the monolayer Fermi surfaces were not located near the $M$-points in those works, and so the effects we describe are significantly different \cite{sinha2021twisting, crasto2019double, wan2025higher}. Another recent work has considered the symmetries of bilayers with different stacking configurations \cite{perkins2025symmetry}.}
\footnotetext[2]{Additionally, the anisotropic band-flattening we shall demonstrate to be induced by the moir\'e potential suppresses Fermi-surface nesting, and is therefore expected to reduce the tendency towards CDW order.}
\footnotetext[3]{On the kagome lattice, the Bloch wavefunctions have non-trivial sublattice structure encoded by  $u_{\mathbf{p}',\sigma'}$; depending on whether the van Hove singularity is `$m$-type' or `$p$-type', $u_{\mathbf{p}',\sigma'}$ has support on either one or two sublattices only (see the SM).}
\footnotetext[4]{Previous studies \cite{kariyado2019flat, kariyado2023twisted} have argued that quasi-one-dimensional physics can arise in $M$-valley bilayers as a result of a so-called `destructive interference manifold', which was claimed to force the tunneling potential to vanish along a line. By construction, Eq. \eqref{tunnel2d} is manifestly two-dimensional, i.e. we find that beyond the leading harmonic, two dimensional harmonics are not forbidden by symmetry. Rather, a quasi-one-dimensional moir\'e potential arises from the momentum dependence of the tunneling amplitude, which results in $w_{1}\gg w_{2}$. In Refs. \cite{jiang20242d, cualuguaru2024new}, ab initio calculations of twisted bilayer 1T-ZrS$_2$ and 1T-SnSe$_2$ highlight that depending on the orbital structure of the active monolayer bands, the moir\'e bands can be two- or one-dimensional. In 1T-ZrS$_2$ for example, the monolayer dispersion $\varepsilon_2 \approx k_x^2/2m_x + k_y^2/2m_y$ with $m_y \gg m_x$ implies the active monolayer states disperse more strongly parallel to $\mathbf{M}_2$, relatively enhancing $w_2$ and producing two-dimensional moir\'e bands. By contrast, the disparity between $m_{x,y}$ is not so significant in 1T-SnSe$_2$, allowing for quasi-one-dimensional bandstructure.}
\footnotetext[5]{If, instead, the residual point group were $C_{2v}$, $s_x$ would be replaced by $s_y$ in \eqref{SOCTunneling}.}
\footnotetext[6]{The number of band-folded states is also regularised by considering momenta sufficiently far away from the vHS that the monolayer dispersion no longer can be approximated as a saddle point.}
\footnotetext[7]{In addition to the 135 kagome metals, 166 variants recently  been studied which also exhibit vHS near the Fermi level and CDW phases, within which the bandstructure is quasi-two-dimensional \cite{arachchige2022charge, korshunov2023softening, tan2023abundant, pokharel2023frustrated, cao2023competing, ortiz2025stability, wang2025formation}. Given these materials are much harder to cleave, we consider them less natural candidate materials for heterostructures.}


\title{Moir\'e $M$-valley bilayers: quasi-one-dimensional physics, \\ unconventional spin textures and twisted van Hove singularities}


\author{Julian Ingham}
\email{ji2322@columbia.edu}
\affiliation{Department of Physics, Columbia University, New York, NY, 10027, USA}

\author{Mathias S.~Scheurer}
\affiliation{Institute for Theoretical Physics III, University of Stuttgart, 70550 Stuttgart, Germany}

\author{Harley D.~Scammell}
\affiliation{School of Mathematical and Physical Sciences, University of Technology Sydney, Ultimo, NSW 2007, Australia}

\begin{abstract}

Motivated by the discovery of quasi-two-dimensional kagome metals AV$_3$Sb$_5$, we consider the theory of twisted bilayers in which the Fermi surface is near the $M$-point. Surprisingly, unlike twisted bilayers of graphene or transition metal dichalcogenides, the moir\'e potential is quasi-one-dimensional: at each $M$-valley, the potential flattens the dispersion strongly along one direction, and weakly along the perpendicular direction. The combination of spin-orbit coupling and twist-induced broken inversion symmetry results in a similarly anisotropic `\textit{moir\'e-Rashba}' potential, which spin-splits the dispersion into coexisting two-dimensional and quasi-one-dimensional bands. We discuss novel aspects of the interplay between mixed dimensionality and spin textures in this platform. First, an applied electric field produces spin polarisation which can be tuned by doping, suggesting potential spintronics applications. Secondly, an in-plane magnetic field momentum- and spin-polarises the Fermi surfaces, producing unconventional spin density waves. Thirdly, in the small-twist-angle limit, the large density of states due to a twisted van Hove singularity near $M$ results in a dense energy spectrum. Our results demonstrate a new variation of moir\'e bandstructure engineering, instigating the study of spin-textured one-dimensional physics in moir\'e materials.

\end{abstract}

\maketitle

\section{Introduction}
Recent years have seen major advances in the fabrication of heterostructures consisting of twisted stackings of two-dimensional materials, most famously twisted bilayer graphene. The tunneling between the twisted layers produces a so-called moir\'e potential which is periodic on a scale much larger than the monolayer lattice constant, folding the monolayer dispersion into a mini Brillouin zone and forming mini bands with a reduced bandwidth \cite{lopes2007graphene, lopes2012continuum, bistritzer2011moire, Khalaf2019}, producing novel interaction physics. In graphene, the chemical potential intersects the bands near the $K$-point of the Brillouin zone, while later studies have considered moir\'e transition metal dichalcogenides (TMDs) with pockets near the $K$ \cite{wu2019topological, yu2020giant, devakul2021magic, zhang2024polarization, jia2024moire, li2021lattice, naik2018ultraflatbands} or $\Gamma$-point, providing a realisation of the hexagaonal lattice Hubbard models \cite{angeli2021gamma}.

Another class of well-studied materials is the recently discovered family AV$_3$Sb$_5$ (A=Rb,K,Cs), which have attracted immense interest due to their demonstration of a rich array of electronic phase transitions \cite{Ortiz2019, Ortiz2020, Ortiz2021, Kang2021, Li2021b, Zhao2021, Li2021c, Mielke2021b, Chen2021, Liang2021, Denner2021, Scammell2023, ingham2025vestigial, Wu2021, Consiglio2021, Jiang2021, hossain2025field}. The materials consist of a stack of weakly bonded exfoliable layers: a triangular lattice of the alkali metal A, a honeycomb lattice of antimony, and a layer consisting of a kagome network of vanadium interposed with a triangular lattice of antimony. Experimental studies have demonstrated these materials can be reduced to the few-layer limit \cite{monokag,song2023anomalous}. Unlike previously studied TMDs or graphene, these systems feature hexagonal Fermi surfaces with van Hove singularities (vHS); the density of states is predominantly located near the $M$-point, rather than $K$ or $\Gamma$.

\begin{figure*}[t]
\centering
	\includegraphics[width = 0.9\textwidth]{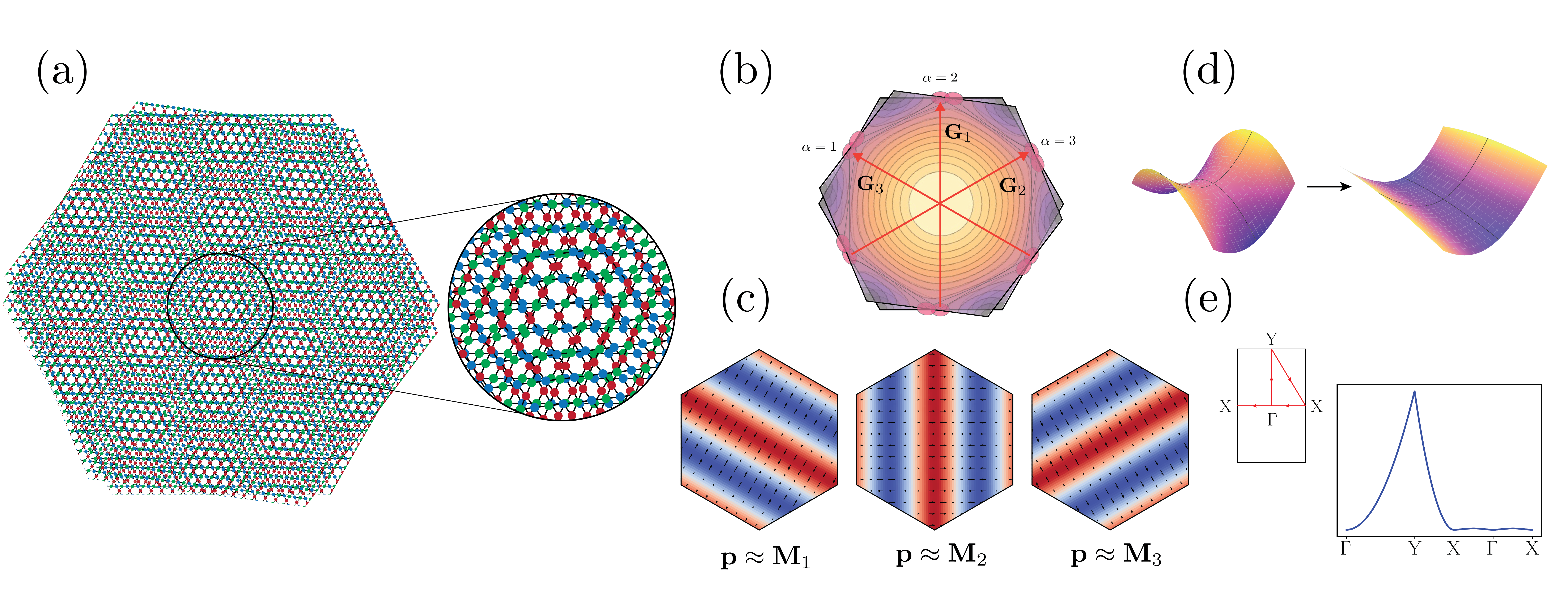}		
	\caption{\textbf{Twisted bilayers of $M$-point Fermi surfaces with van Hove singularities.} (a) Twisted sheets of kagome monolayers. (b) The twisted Brillouin zones featuring hexagonal Fermi surfaces with corners located at the $\mathbf{M}_\alpha$ points, where the density of state diverges. (d) Each distinct $M$-point experiences a quasi-one-dimensional moir\'e potential, resulting in a preferential flattening of the saddle point dispersion along one direction; moir\'e bandstructure without spin-orbit centered at a moir\'e $\Gamma$ point near $\mathbf{M}_2$ is shown in (e). Twist-angle-induced inversion symmetry breaking produces a similarly one dimensional spin orbit potential at each $M$-point; the spin-dependent tunneling potential near each $M$ point is shown in real space in (c).}
	\label{fig1}
    \vspace{-0.3cm}
\end{figure*}


Motivated by the possibility of producing twisted bilayers of kagome metals, we consider the problem of twisted bilayers with Fermi pockets near the $M$-point with spin-orbit coupling. We find that the analysis of twisted bilayers with $M$-point Fermi surfaces differs strikingly from the comprehensively studied problem of twisted graphene and transition metal dichalcogenides: at each of the three vHS, the leading harmonics in the moir\'e superlattice potential are \textit{one dimensional} -- i.e. they consist of a single wavevector -- and twist-induced broken inversion symmetry results in similarly one-dimensional, emergent spin-orbit coupling -- which we refer to as \textit{moir\'e Rashba} coupling. Depending on the monolayer bands, the result can be bands which flatten anisotropically, resulting in quasi-one-dimensional bands with novel spin textures.

In this paper we discuss a number of novel features of these bands. The large density of states associated to a vHS results in a large number of Bloch states band-folded on top of each other in the presence of a superlattice potential, behaviour which is regularised at larger twist angles. Firstly, in the larger twist angle regime, the moir\'e-Rashba coupling spin-splits the Fermi surfaces into bands with distinct spin textures; we find that by tuning the strength of spin-orbit and interlayer tunneling, it is possible to engineer a novel situation in which the Fermi surface consists of two kinds of contour -- one hosting weakly-confined, delocalised two-dimensional excitations, and the other hosting strongly confined, quasi-one-dimensional excitations situated near the minima of the moir\'e potential. Second, the lowest energy bands form strongly spin-momentum-locked valleys, so that an applied electric field produces a spin current which can be tuned by doping -- suggesting prospects for moir\'e spintronics. Thirdly, an applied in-plane magnetic field causes the one- and two-dimensional bands to become spin polarised, so that one spin species is confined into quasi-one-dimensional channels while the other is delocalised in the two-dimensional bilayer. Fourthly and finally, we present a tentative discussion of the small twist angle limit, in which the moir\'e potential couples a large density of band-folded states, producing a dense spectrum of moir\'e bands.

Our work represents the first study of twisted vHS. Moreover, the quasi-one-dimensional tunneling and spin-orbit coupling are neither limited to kagome layers, nor to systems with vHS, but apply generally to twisted bilayers of materials whose Fermi surfaces are located at momenta in the Brillouin zone with a $C_{2}$ little group, and stimulate new directions in the study of moir\'e heterostructures.

\section{Results}

\subsection{Continuum model for $M$-valley bilayers}

We begin by deriving a general model of a twisted bilayer with Fermi surfaces near the $M$-point, or $M$-valleys. We work with a theory in which the full Fermi surface is truncated to the near vicinity of the $M$-point -- referred to as `patches'. Since opposite corners are related by a reciprocal lattice vector, there are three inequivalent patches, which are indexed by $\alpha=1,2,3$, associated to the momenta $\mathbf{M}_\alpha$.

In the case of kagome metals, bulk AV$_3$Sb$_5$ possesses vHS near the $M$-point, so that the density of states is peaked in this region, resulting in a hexagonal Fermi surface as shown in Fig. \ref{fig1}b. Considering a monolayer with a similar dispersion would justify a patch model, but we may also imagine a more prosaic scenario in which a monolayer dispersion simply possesses electron or hole pockets near $M$ \cite{Note1}. Kagome metals are known to develop charge density wave (CDW) order at $T_{\text{CDW}}\sim 90 $ K, but first principles calculations indicate the interlayer tunneling may be of the order $t_\perp \sim$ eV, in which case the high-temperature normal state of the twisted bilayer may be considered without incorporating the CDW\cite{Note2}.

The general bilayer Hamiltonian $\mathcal{H} = \mathcal{H}_\parallel + \mathcal{H}_\perp$ is 
\begin{align}
    \mathcal{H}_\parallel &= \sum_{i, a}\int d\mathbf{r}d\mathbf{r}' \ t(\mathbf{r}-\mathbf{r}')c^\dag_{i}(\mathbf{r})c_{i}(\mathbf{r}') \\
    \mathcal{H}_\perp &=\!\! \sum_{ab}\int d\mathbf{r}d\mathbf{r'} \ c^\dag_{i}(\mathbf{r}) T^{ij}(\mathbf{r},\mathbf{r}')c_{j}(\mathbf{r}')
\end{align}
where $i$ indexes layer, $t(\mathbf{r}-\mathbf{r}')$ is the intralayer hopping and $T^{ij}(\mathbf{r},\mathbf{r}')$ the interlayer tunneling. We treat the interlayer tunneling in the two-center approximation; our conclusions may be modified when interlayer assisted hopping is important. We assume tunneling between a single band in each layer, but present more general expressions in the Supplementary Material (SM). We work in an untwisted reference coordinate system, in which the locations of the atoms in layers $i$ and $j$ are rotated by $\theta_i$ and $\theta_j$. 

Under the assumption that the interlayer tunneling is translationally and rotationally invariant on the moir\'e scale, we may rewrite the tunneling potential in momentum space (see SM Sec. \ref{cont_deriv}),
\begin{gather}
    \mathcal{H}_\perp =\sum_{\sigma\sigma',\mathbf{p}\mathbf{p}',\mathbf{G}\mathbf{G}'}  T^{ij}(|\mathbf{p}+\mathbf{G}|)e^{i\mathbf G \cdot \mathbf{r}_\sigma}e^{-i\mathbf{G}' \cdot \mathbf{r}_{\sigma'}} \nonumber \\
    \times \ u_{\mathbf{p}',\sigma'} u_{\mathbf{p},\sigma}\psi^\dag_{i,\mathbf{p}}\psi_{j,\mathbf{p}'}
\label{moirepot1}
\end{gather}
where $\sigma$ indexes sublattice,  $u_{\mathbf{p}',\sigma'}$ is a Bloch function encoding the sublattice dependence of the orbital wavefunctions, and we have the restriction that
\begin{align}
\label{pcons}
    R_{\theta_i}(\mathbf{p}+\mathbf{G})=R_{\theta_j}(\mathbf{p}'+\mathbf{G}')
\end{align}
where $R_{\theta_i}$ is a rotation matrix and $\mathbf{G}$ and $\mathbf{G}'$ are (unrotated) monolayer reciprocal lattice vectors \cite{Note3}. We can understand these expressions physically: a free electron with momentum $R_{\theta_i}(\mathbf{p}+\mathbf{G})$ in layer $i$ may scatter to a state $R_{\theta_j}(\mathbf{p}'+\mathbf{G}')$ in layer $j$, while the tunneling term contains a relative phase factor when the initial and final states are associated to different sublattices.

Taking $\mathbf{p}$ and $\mathbf{p}'$ as close to the $M$-points $\mathbf{M}_\alpha$ and $\mathbf{M}_{\alpha'}$, satisfying \eqref{pcons} requires $\alpha=\alpha'$ and $\mathbf{G}=\mathbf{G}'$ for small twist angles. The tunneling amplitude $T(|\mathbf{q}|)$ is generically expected to fall off rapidly for large momenta, so the dominant contributions to the moir\'e potential arise from $\mathbf{p}+\mathbf{G}$ located in the first Brillouin zone. As shown in Fig. \ref{fig1}b, the only two momenta $\mathbf{G}$ such that $\mathbf{p}+\mathbf{G}$ stays in the first Brillouin zone are $\mathbf{G}^{(1)}_n=0, -2\mathbf{M}_\alpha$. Hence, the moir\'e potential in the leading approximation only contains two possible momentum transfers (Fig. \ref{fig1}b). The two next-leading harmonics in the moir\'e potential correspond to $\mathbf{G}^{(2)}_m=-2\mathbf{M}_{\beta}$ with $\beta\neq\alpha$, producing a weak two-dimensional profile in the moir\'e potential. 

Assuming that we may treat the orbital overlaps and the tunneling amplitude as roughly constant in the vicinity of the patches, the sublattice and orbital dependence can be absorbed into an overall constant prefactor (see SM Sec. \ref{cont_deriv}). After Fourier transforming the tunneling matrix elements in Eq.~\eqref{moirepot1} for patch $\alpha$ back to real space,
\begin{gather}
T^{ij}(\mathbf{r}) = ( w_{1}\sum_{n} e^{i\mathscr{R}(\mathbf{G}^{(1)}_n+\mathbf{M}_\alpha)\cdot \mathbf{r}}  \nonumber \\
\label{tunnel2d}
+w_{2}\sum_{m} e^{i\mathscr{R}'(\mathbf{G}^{(2)}_m+\mathbf{M}_\alpha)\cdot \mathbf{r}} ) (\ell_x)_{ij} \\
=\left(w_{1}\cos(\mathscr{R}\mathbf{M}_\alpha\cdot \mathbf{r})+w_{2}\cos(\sqrt{3} \mathscr{R}'\mathbf{M}_\alpha\cdot \mathbf{r}) \right) (\ell_x)_{ij}\nonumber
\end{gather}
where $\mathscr{R}=R_{\theta_i}-R_{\theta_j}$, $\mathscr{R}'=R_{\theta_i+\pi/2}-R_{\theta_j+\pi/2}$, and $\ell_x$ is a Pauli matrix in layer space. This expression is the real space form of the moir\'e potential experienced by electrons in the vicinity of $\mathbf{M}_\alpha$. Neglecting the subleading harmonic $w_{2}\ll w_1$, one finds that the moir\'e potential consists of a one-dimensional modulation $\propto \cos(\mathscr{R}\mathbf{M}_\alpha\cdot \mathbf{r})$ \cite{Note4}. In the bandstructure plots which follow, we shall neglect the subleading harmonic associated to $w_{2}$, but discuss its effects further in the SM and Sec. \ref{twist_vhs}. We caveat that when the monolayer orbitals have strong spatial structure, with a larger dispersion parallel to $\mathbf{M}_\alpha$ than perpendicular, then $w_2$ can become comparable to $w_1$, as discussed in Ref. \cite{cualuguaru2024new}.


Hence, the only momentum-conserving tunneling processes at long wavelengths are those which couple $\mathbf{M}_\alpha$ and $-\mathbf{M}_\alpha$, which may be contrasted with the situation in twisted bilayer graphene. For a Fermi surface near the $K$-point, the lowest-order momentum transfers have components in both the $k_x$ and $k_y$ directions, which is related to the fact there are three equivalent $K$-points in the Brillouin zone. In the present case, the dispersion in the vicinity of a patch is flattened on one direction, and left untouched in the perpendicular direction; in the limit where the moir\'e potential flattens the bands completely, the result will be quasiparticles which are dispersive along only one direction in momentum space (Fig. \ref{fig1}e). 

\begin{figure*}[t!]
    \centering
    \includegraphics[width=\textwidth]{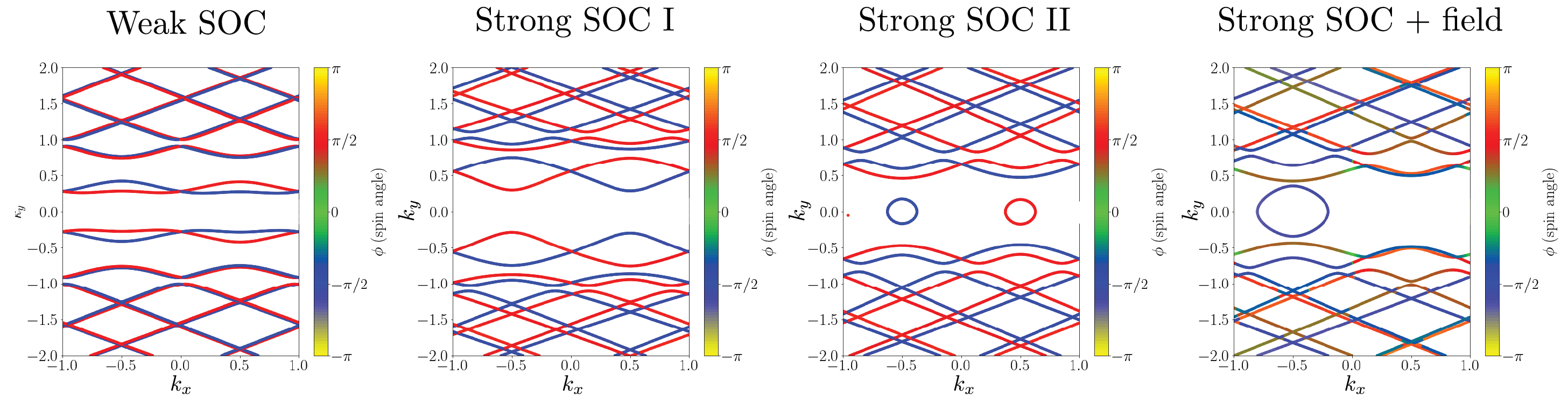}
    \vspace{-0.7cm}
    \caption{\textbf{Continuum model band structure with spin-orbit coupling and applied Zeeman field.}  Energy contours with $w_1=0.5$, $w_2=0$; spin angle $\phi = \tan^{-1}(s_y/s_x)$ shown in colour. Units of energy $tQ^2$ and momentum $Q$ are set to unity, and the twist angle is taken to be $\theta=1.05^\circ$.  (a) Weak spin-orbit: $v_1=u_1=0.05 w_1$ with  $E_F=-0.45$. (b) and (c) Strong spin-orbit:  $v_1=u_1=$ $0.05 w_1$, with (b) $E_F=-0.3$ and  (c) $E_F=-0.7$. (d)  Strong spin-orbit and strong field: $v_1=u_1=0.25 w_1$,  $B_y=0.125$ and $E_F=-0.7$. Plots are shown for the vicinity of $\mathbf{M}_2$; the dispersion at $\mathbf{M}_{1,3}$ is related by threefold rotation.}
    \label{fig:cont_disp}
    \vspace{-0.3cm}
\end{figure*}

\subsection{Moir\'e-Rashba potential} 
Three-dimensional inversion symmetry is broken by the twist angle; when the monolayers involve heavier elements, spin-orbit coupling becomes important, contributing a \textit{moir\'e-Rashba} potential. If the individual monolayers have inversion symmetry, the bands are still pseudospin degenerate in the absence of tunneling between the layers (we index pseudospin via Pauli matrices $s_i$ in what follows). Hence, to leading order, the pseudo-spin splitting is generated by interlayer tunneling terms. Given the arguments above that the interlayer tunneling only involves the momentum transfer $\pm \mathbf{Q}_\alpha$, one can deduce the general form of the spin-orbit coupling on symmetry grounds. Taking the residual point group at a single $M$-point to be $D_2$, as is the case in kagome metals \cite{Note5}, we find spin-orbit contributes a moir\'e potential 
\begin{gather}
\label{SOCTunneling}
    \mathscr{S}(\mathbf{r}) = v_{1}\sin(\mathscr{R}\mathbf{M}_2\cdot \mathbf{r}) \ell_y s_y + u_{1}\cos(\mathscr{R}\mathbf{M}_2\cdot \mathbf{r}) \ell_y s_z 
\end{gather}
for patch $\alpha=2$ (next-leading harmonics are given in SM Sec. \ref{supp-rashba}); the spin-orbit potential at the other patches are related by threefold rotation. The moir\'e spin-orbit potential effectively corresponds to a co-planar spiral spin texture, spatially modulated with the moir\'e period, see Fig. \ref{fig1}c. 
We contrast these effects with spin-orbit coupling in moir\'e TMDs, in which the monolayers already break inversion symmetry and have significant spin-splitting as a result \cite{kormanyos2015k}. The moir\'e-Rashba potential we describe here is a qualitatively different effect, specifically arising due to the effect of the twist angle; as opposed to TMDs it is not only highly non-collinear but also breaks translation symmetry, and hence couples different momenta. 

Incorporating spin-orbit, we arrive at the continuum model Hamiltonian, which we write for $\alpha=2$ as
\begin{gather}
\mathcal{H} = \sum_{\mathbf{k},\tau} \psi^\dag_{2,\mathbf{k}} \varepsilon_{2}(R_{\ell_z \theta/2}\mathbf{k})\psi_{2,\mathbf{k}}
\!+ w_1 \,\psi^\dag_{2,\mathbf{k}} \ell_xs_0\psi_{2,\mathbf{k}+\tau\mathbf{Q}_2} \nonumber \\
+ \tau v_1\, \psi^\dag_{2,\mathbf{k}} \ell_x s_y\psi_{2,\mathbf{k}+\tau \mathbf{Q}_2} + u_1\, \psi^\dag_{2,\mathbf{k}} \ell_y s_z \psi_{2,\mathbf{k}+\tau \mathbf{Q}_2} 
\label{continuum_model_main}
\end{gather}
where $\mathbf{Q}_2\equiv\mathscr{R}\mathbf{M}_2=(4\pi/a\sqrt{3})(\sin(\theta/2),0)$, $\mathbf{k}$ is momentum relative to the $M$-point, and $\psi^\dag_{\alpha,\mathbf{k}}$ is a four-component creation operator in layer and spin space describing an excitation within the close vicinity of $\mathbf{M}_\alpha$. We refer to $\tau=\pm$ as valleys. In our bandstructure results, we shall focus on $\alpha=2$, with the understanding that the dispersion at the other patches is related by threefold rotation. 

\subsection{Twisting van Hove singularities}
\label{twist_vhs}

Motivated by kagome metals, we shall consider a saddle-like monolayer dispersion $\varepsilon_{\alpha}(\mathbf{k})$ near $\mathbf{M}_\alpha$ arising from the vHS of a hexagonal Fermi surface, with the form e.g. $\varepsilon_{2}(\mathbf{k}) = \tfrac{1}{4}t(-k_x^2+3k_y^2)$. Our preceding analysis relies predominantly on the location of the Fermi surface and symmetries of the problem; below we present the first discussion of twisted vHS.

Consider first the zero twist angle limit. A saddle point has flat directions -- there are an infinite number of plane wave states $k_x=\pm \sqrt{3} k_y$ which have distinct momenta yet zero energy. This is quite unlike the ordinary case of a monotonic band dispersion, in which the states which get band folded into the first Brillouin zone all have different energies, e.g. $(\mathbf{k}-n\mathbf{Q})^2/2m$. Since for a vHS a large number of states have \textit{zero} energy, the large density of states at a vHS is converted into a large number of moir\'e bands, which become dense as one includes states further and further away from the saddle point.

To illustrate, consider the effects of the first harmonic $\mathbf{Q}_2$; setting the Fermi energy to zero, states from the $n$th Brillouin zone contribute to the Fermi surface when
\begin{align}
    \varepsilon_2(\mathbf k + n \mathbf Q_2)&=\tfrac{1}{4}t(-(k_x + n Q_{2x})^2+3 k_y^2) =0
\end{align}
Solutions exist for all $n$, producing the criss-crossed Fermi surfaces at increasingly large $|k_y|$ in Fig. \ref{fig:cont_disp}.  Inclusion of the next-leading harmonic $\mathbf{Q}'_2=\mathscr{R}'\mathbf{M}_2$ folds this infinite set of criss-crossing bands to near the origin. Since 
$\varepsilon_2(n \mathbf Q_2 + m \mathbf{ Q}'_2) = 0$ for $|n|=3|m|$, there are an infinite number of states back-folded. The behaviour is regularised with increasing twist angles, as the flat directions of the monolayer dispersions become far enough separated that there is an increasingly fewer number of band-folded states \cite{Note6}.

In the limit of weak superlattice potential, the bands which overlap from band-folding do not mix at single-particle level, and these bands are essentially decoupled single-particle states of the monolayer folded to zero momentum. At larger superlattice strength, these dense bands hybridize; the avoided crossings opened up by the crossing of this large number of bands can introduce very rapid changes in phase of the eigenstates as one varies the energy in a narrow range. The combination of a large number of near-degenerate ``flavour'' degrees of freedom with eigenstates whose relative phases vary rapidly may produce interference processes upon the inclusion of interactions which resemble the physics of the SYK model \cite{sachdev1993gapless, maldacena2016remarks}; an accompanying paper will explore this aspect more precisely.
 
 \subsection{Aspects of the bandstructure}

\subsubsection{Spin-valley locking}
\label{svl}

The moir\'e-Rashba potential introduces a spin-dependent tunneling term, resulting in moir\'e bands with interesting spin textures.  From Fig. \ref{fig:cont_disp}, one sees in the absence of external field that the bands with positive and negative momenta are spin polarised with opposite spins. In Fig. \ref{fig:cont_disp}a and b, the splitting appears vanishingly small at $k_x=0$.  Depending on the Fermi energy, the bands either consist of small two-dimensional pockets with nearly complete spin polarisation in one direction locked to the momentum, or quasi-one-dimensional bands in which both spin textures are present.

To understand these features, we state the effective Hamiltonian for the spin-split bands near the $\mathbf{M}_2$ point, starting from the saddle point dispersion $\varepsilon_{2}(\mathbf k)$ and treating $w_1$,$v_1$,$u_1$, perturbatively (see SM Sec. \ref{supp-heff}), 
\begin{gather}
{\cal H}_{k\cdot p} \\
=\varepsilon_2(R_{\ell \theta/2}\mathbf k)+ (\alpha_1 k_x + \alpha_2 \ell k_y)(\gamma_1 s_x +  \gamma_2 \ell s_y) - {\mathbf B} \cdot {\mathbf s} \nonumber
\label{main-Hkp}
\end{gather}
where we have included a Zeeman field $\mathbf B$ in units $g\mu_B=1$. Here $\gamma_2 =2 w_1 v_1$ and  $\gamma_1 =2 v_1 u_1$, and since we expect $w_1>v_1\approx u_1$  one finds $\gamma_2>\gamma_1$. The near perfect $s_y$ polarization observed in Figs. \ref{fig:cont_disp}a,b\&c can be understood as  $\gamma_2\gg\gamma_1$.  One can deduce from ${\cal H}_{k\cdot p}$ that at each valley $(k_x,k_y)=(\pm Q/2,0)$, the spin is oppositely polarised. Moreover, we find that $\alpha_2/\alpha_1=2\sin\theta/(2\cos\theta -1)$ and hence at small twist angle spin-splitting due to $\alpha_1$ is more significant than due to $\alpha_2$; considering $k_x=0$ and $k_y\neq0$, this accounts for the vanishingly small splitting of Fig. \ref{fig:cont_disp} as well as the non-zero splitting observed for larger twist angle (see SM). 
The unusual dependence of \eqref{main-Hkp} on $\mathbf{k}$, and the spin texture shown in Fig.~\ref{fig:cont_disp}, owes to the peculiar realisation of $C_{2x}$ symmetry on the moir\'e bands, which acts non locally in momentum space (see SM Sec. \ref{syms}).

One consequence of the perfect spin-valley locking is that an applied electric field in the $+x$ direction results in a net spin polarised in the $-x$ direction. More quantitatively, one may see this in a  Boltzmann equation treatment. In the presence of an applied electric field, the Fermi distribution function is shifted to the equilibrium distribution $f^{(0)}(E_{s\mathbf{k}})$ plus $g_s(\mathbf{k})$; solution of the Boltzmann equation results in 
\begin{align}
    g_s(\mathbf{k}) = -e \mathbf{E}\cdot \mathbf{k} \frac{\hbar \tau_0}{m^*}\left[-\frac{\partial f^{(0)}(E_{s\mathbf{k}})}{\partial E_{s\mathbf{k}}} \right]
\end{align}
where we have assumed a static electric field and an energy- and momentum-independent scattering time $\tau_0$. Note that here $s$ denotes the $y$-projection of the spin, which is a good quantum number for the spin-valley locked Fermi surfaces. In the low temperature limit, $\partial f^{(0)}/\partial E_{s\mathbf{k}} \approx -\delta (E-E_{s\mathbf{k}})$. Taking the electric field to be along the $y$ direction, 
\begin{align}
    \langle s_x \rangle &= \sum_s \int \frac{d^2 k}{(2\pi)^2} s g_s(\mathbf{k}) \nonumber \\
    &\approx - \frac{eE \hbar \tau_0}{m^*} \sum_s \int \frac{d^2 k}{(2\pi)^2} s k_y \delta (E - E_{s\mathbf{k}}) \neq 0
\end{align}
The other two pairs of valleys at $\pm \mathbf{Q}_1/2$ and $\pm \mathbf{Q}_3/2$ each give a smaller contribution to the spin polarisation by a factor $\cos(2\pi/3)=1/2$.


This situation is similar to the persistent spin helix studied in semiconductor 2DEGs, which has been used to create stable spin textures with an applied current \cite{Bernevig2006, koralek2009emergence}.  In other 2D materials with spin-valley locking such as transition metal dichalcogenides -- in which spin-charge conversion has been observed \cite{ghiasi2019charge} -- the bands at each valley feature both spins, but with an imbalanced population. By contrast, the moir\'e valleys here are totally spin-valley polarised, which may enhance the spin-charge conversion. On the other hand, the associated carrier density will be small in the small twist angle limit, which may reduce the effect. The possible use of $M$-valley moir\'e heterostructures for producing electrically tunable spin polarisation motivates experimental investigation of this system for spintronics applications.

\subsubsection{Unconventional Zeeman splitting}
\label{zeeman}

Including a magnetic field induces spin and momentum polarisation. This effect is most easily seen in the regime (here tuned via the Fermi energy) of closed 2D moir\'e Fermi surfaces, as opposed to the open, quasi-1D Fermi surfaces. As such, Fig. \ref{fig:cont_disp}d presents a spin and momentum polarisation at a Fermi energy corresponding to a closed Fermi surface for the moir\'e bands. 


For appropriately chosen parameters, one can create a situation in which one spin species is confined into quasi-1D channels, while the other is unconfined, producing coexisting quasi-1D and 2D spin-polarised quasiparticles. Fig. \ref{fig:cont_disp}(d) presents this regime, meanwhile in the SM we show a transition of the spin-polarised Fermi surfaces as a function of chemical potential:  First, there is a quasi-1D spin-polarised band, then there is both quasi-1D and 2D bands, with opposite spin polarization; finally both spin species are 2D.

The direction of the confined spin projection varies from patch to patch by threefold rotation; in real space, this manifests as three sets of parallel quasi-one-dimensional channels, relatively rotated by $2\pi/3$. Twist angle can be tuned to interpolate between weak and strong quasi-one-dimensional confinement, allowing experiments to explore the crossover between these two regimes. \vspace{-0.3cm}

\begin{figure}[t!]
    \centering
	\includegraphics[width = 0.43\textwidth]{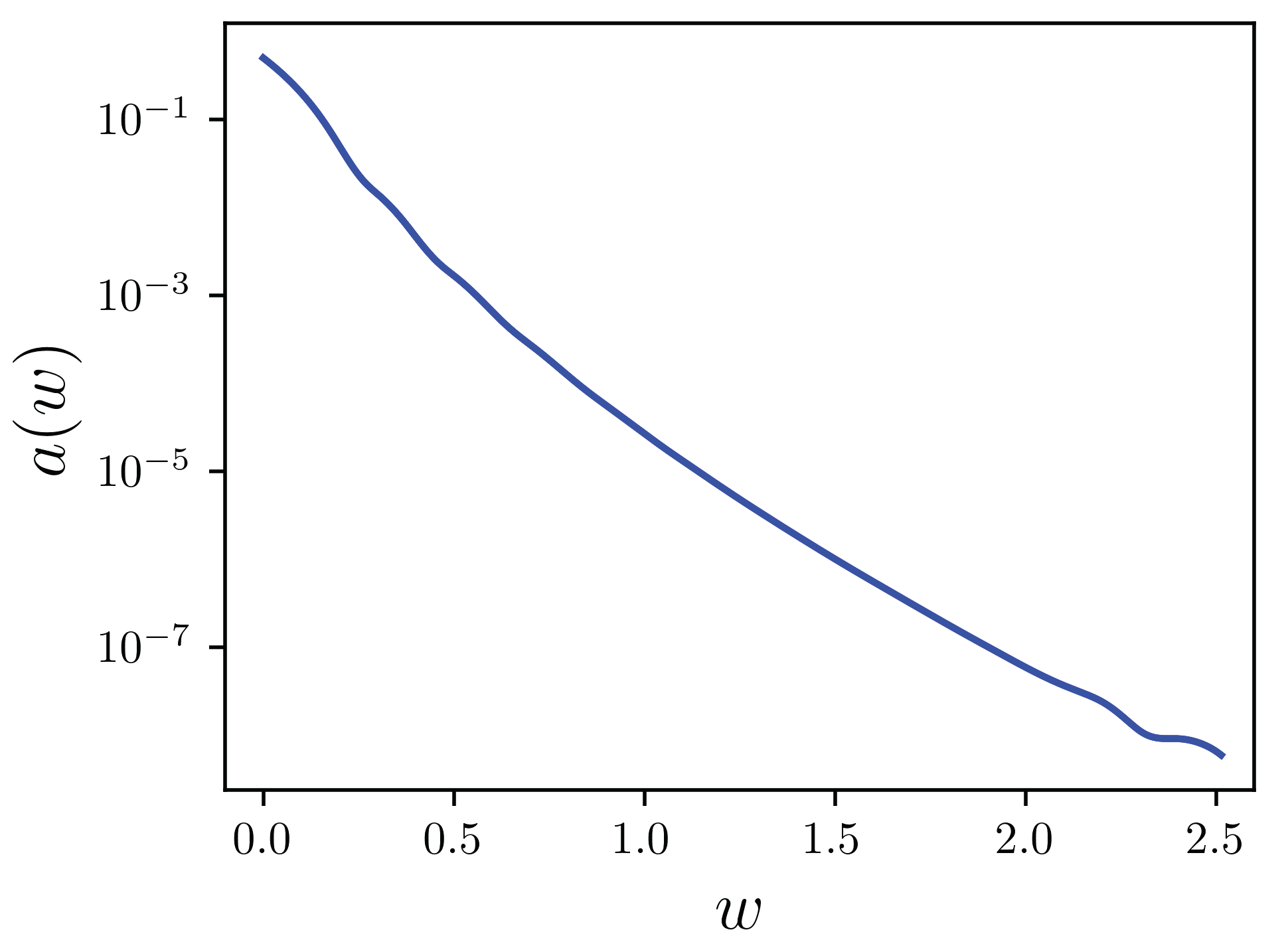}	

    \vspace{-0.3cm}
	\caption{\textbf{Anisotropic effective mass enhancement.} Inverse effective mass perpendicular to $\mathbf{M}_2$ in the single-harmonic model, defined so $E(\mathbf{k}) = a(w)k_x^2 - \tfrac{3}{2}k_y^2$.}
	\label{fig:a_param}
    \vspace{-0.5cm}
 \end{figure}
 
\subsubsection{Anisotropic band flattening}
\label{1dflat}

In the one-dimensional limit,  the bands flatten predominantly along the direction of the leading harmonic; defining $E(\mathbf{k}) = a(w_1)k_x^2 - \tfrac{3}{2}k_y^2$ near $\mathbf{M}_2$, we plot the suppression of the $k_x$ inverse effective mass in Fig. \ref{fig:a_param}; the monolayer corresponds to $a(0)=1/2$. One finds the bands are effectively one-dimensional for relatively small values of $w_1$. The suppression of all derivatives of the dispersion along $k_x$ generates a ``higher-order'' vHS \cite{Shtyk2017, Chandrasekaran2020, Classen2020, Chandrasekaran2023a, chandrasekaran2024engineering, Han2023, nag2024pomeranchuk, wang2024classification}; it is curious to note that the one-dimensional limit is in some sense an ``infinite order'' van Hove point.

\section{Discussion}

The twisted $M$-valley bilayer bandstructure produces emergent quasi-one-dimensional bandstructure, with unconventional spin textures depending on the strength of spin-orbit coupling. Possible experimental probes of the quasi-one-dimensional normal state include transport anisotropy, gate-tunable spin current, and the presence of a quasi-one-dimensional local density of states probed by STM with and without spin resolution. 

While we have used a toy model of V$_3$Sb as a motivating example \cite{Note7}, aspects of the theory we have described are simply reliant on the $M$-point Fermi surface, or spin-orbit coupling, rather than the specifics of the kagome lattice or the presence of a vHS. Other candidate two-dimensional materials -- i.e. those which have Fermi pockets at the $M$-point, or $X$-point in a $C_4$ symmetric system -- may be a better realisation of the general phenomena. Some possible candidates include monolayer Mg$_3$Sb$_2$ \cite{huang2019significant}, Cu-benzenehexathial \cite{zhang2017theoretical}, FeSe \cite{si2023iron}, NbSe$_2$ proximitised by two-dimensional ferromagnet CrBr$_3$ \cite{kezilebieke2020topological}, and certain group IV-B transition metal dichalcogendies such as monolayer TiSe$_2$ and HfTe$_2$ \cite{lasek2021synthesis,tsipas2021epitaxial}. Note however that while $w_1>w_2$ is expected for isotropic monolayer orbitals, strong one-dimensionality depends on the regime $w_1\gg w_2$ -- providing a guiding criterion in bandstructure analyses. To realise the regime of mixed-dimensionality, ab initio studies should search for materials with larger spin-orbit coupling. Candidate materials may potentially be explored within high-throughput searches of exfoliable materials, e.g. Ref \cite{Mounet2018, jiang20242d}. Quasi-one-dimensional physics has been observed in twisted WTe$_2$ \cite{Wang2022}, though the one-dimensional channels in this system arise due to different effects; our results suggest a systematic approach to finding other moir\'e heterostructure hosting novel quasi-one-dimensional physics.

The general effect -- of anisotropic band flattening for small-twist hexagonal Fermi surfaces -- might potentially also be realised in a layered bulk compound, in which adjacent weakly-coupled layers along the $c$-axis are relatively misaligned, potentially due to disorder. This effect may possibly be related to the anisotropic quasiparticle scattering signature seen in STM studies of CsV$_3$Sb$_5$ \cite{Zhao2021}. Future experiments may consider our model as a possible explanation of such effects as those we have discussed.

One-dimensional systems generically host strong interaction effects, including non-Fermi liquid behaviour and novel ground states \cite{Giamarchi2004} -- a phenomena which has motivated many theoretical studies to consider coupled one-dimensional systems as an avenue to exotic physics in higher dimensions \cite{mukhopadhyay2001sliding, vishwanath2001two, Sondhi2001, Emery2000, Wen1990, Kane2017}. Depending on spin-orbit coupling our proposed system may realise a novel kind of heavy fermion physics, in which two-dimensional itinerant electrons coexist not with point-like local spin moments, but with extended, one-dimensional spin moments. We also make some comments on the simplest case with no spin orbit and no dense spectrum. The patches are essentially decoupled at the single-particle level, as a large momentum transfer is required for a single electron to tunnel between the patches, wowever, since the inter-patch momentum $\mathbf{M}_i$ is half a reciprocal lattice vector it is possible for a pair of electrons to tunnel from one patch to another. Therefore, the allowed four-fermion scattering processes include an intra-patch interaction $g$ and an interpatch umklapp scattering term $v$, resulting in the effective theory
\begin{gather}
\mathcal{H}_{\text{eff}} = \sum_\alpha (E_\alpha(\mathbf{k}) - \mu) \psi^\dag_{\mathbf{k}\alpha}\psi_{\mathbf{k}\alpha} + g \sum_\alpha  \psi^\dag_{\mathbf{k}\alpha}\psi_{\mathbf{k}\alpha}\psi^\dag_{\mathbf{k}\alpha}\psi_{\mathbf{k}\alpha} \nonumber\\
+v  \sum_{\alpha\neq \alpha'}  \psi^\dag_{\mathbf{k}\alpha}\psi_{\mathbf{k}\alpha'}\psi^\dag_{\mathbf{k}\alpha}\psi_{\mathbf{k}\alpha'}
\end{gather}
where $E_\alpha(\mathbf{k})$ is the dispersion of the moir\'e quasiparticles. Hence, our theory produces a set of three flavours of one-dimensional theory coupled by umklapp scattering; future studies may investigate the interaction-induced instabilities of our minimal model. \\

 \textit{Note:} Prior to arXiving our work, Refs. \cite{jiang20242d, cualuguaru2024new, lei2024moir} appeared, which also derive the $M$-valley continuum model and propose candidate materials for quasi-one-dimensional moir\'e bands, focusing in particular on 1T-$MX_2$ ($M$=Zr,Sn and $X$=S,Se). Refs. \cite{jiang20242d, cualuguaru2024new} point out that quasi-one-dimensionality can be understood beyond the leading harmonic or two-center approximations, in terms of the non-symmorphic realisation of mirror $z$ symmetry combined with appropriate real-space orbital structure. Our results are consistent with those references, but also contain significant differences in focussing on a saddle point dispersion rather than the monotonic quadratic dispersion of Refs. \cite{jiang20242d, cualuguaru2024new, lei2024moir}. Our work also introduces moir\'e spin-orbit coupling, which plays a crucial role in many of the effects we describe. 

\begin{acknowledgments}
The authors thank Andrei Bernevig, Madisen Holbrook, Brenden Ortiz, Yves Kwan, Tommy Li, and Raquel Queiroz for discussions and comments. JI is supported by NSF Career Award No. DMR-2340394. M.S.S.~acknowledges funding by the European Union (ERC-2021-STG, Project 101040651---SuperCorr). Views and opinions expressed are however those of the authors only and do not necessarily reflect those of the European Union or the European Research Council Executive Agency. Neither the European Union nor the granting authority can be held responsible for them.
\end{acknowledgments}

\let\oldaddcontentsline\addcontentsline
\renewcommand{\addcontentsline}[3]{}

\bibliography{m_valley_refs.bib}

\onecolumngrid

\newpage

\begin{center}
\textbf{\large Supplementary Material}
\end{center}
\setcounter{equation}{0}
\setcounter{table}{0}
\setcounter{section}{0}
\setcounter{figure}{0}
\renewcommand{\theequation}{S\arabic{equation}}
\renewcommand{\thefigure}{S\arabic{figure}}
\renewcommand{\thesection}{S\arabic{section}}

\section{Derivation of the continuum model}
In this section we derive the continuum model for twisted bilayers with Fermi surfaces at the $M$-point, with specific details for the case of twisted kagome bilayers.

\subsection{Kagome lattice}
\begin{figure}[b]
	\includegraphics[height = 0.35\textwidth]{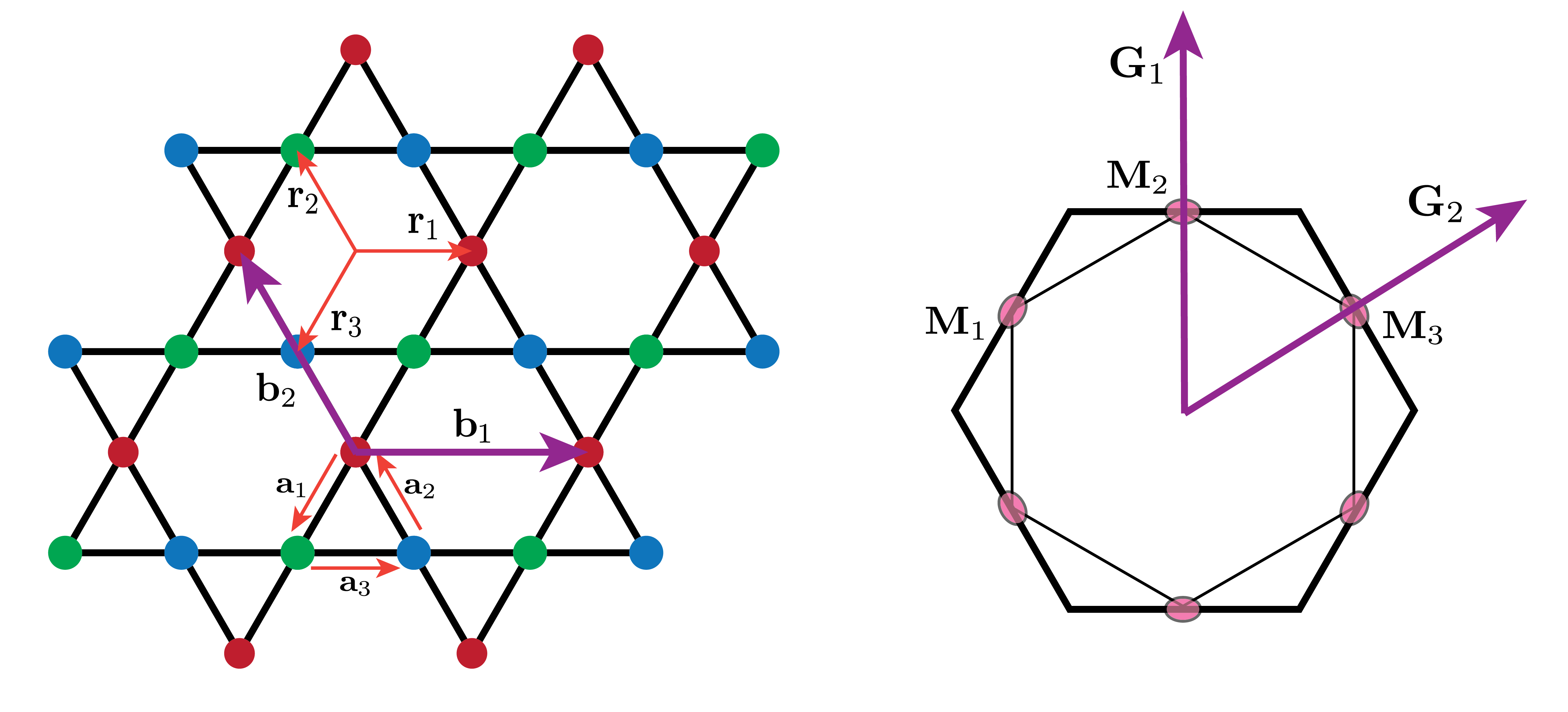}			
	\caption{Left: The kagome lattice with Bravais lattice vectors $\mathbf{b}_2=a(1,0)$ and $\mathbf{b}_1=a(-\tfrac{1}{2},\tfrac{\sqrt{3}}{2})$. Relative to the center of the unit cell, the positions of the $a$ (red), $b$ (green), and $c$ (blue) sublattices are $\mathbf{r}_1=\tfrac{a}{2}(1,0)$, $\mathbf{r}_2=\tfrac{a}{2}(-\tfrac{1}{2},\tfrac{\sqrt{3}}{2})$ and $\mathbf{r}_3=\tfrac{a}{2}(-\tfrac{1}{2},-\tfrac{\sqrt{3}}{2})$.  Right: the Brillouin zone, with reciprocal lattice vectors $\mathbf{G}_1$ and $\mathbf{G}_2$, and patches encircling the $M$-points $\mathbf{M}_1$, $\mathbf{M}_2$, and $\mathbf{M}_3$.}
	\label{fig:kag}
\end{figure}

Before deriving the continuum model for $M$-point Fermi surfaces, we shall first we state some preliminary identities related to the kagome lattice, shown in Fig. \ref{fig:kag} with $a$ (red), $b$ (green), and $c$ (blue) sublattices indicated. 
The Brillouin zone is a hexagon illustrated on the right of Fig. \ref{fig:kag}, with reciprocal lattice vectors are given by 
\begin{align}
\mathbf{G}_1 &= \tfrac{4\pi}{a\sqrt{3}}(\tfrac{\sqrt{3}}{2},\tfrac{1}{2})\\
\mathbf{G}_2 &= \tfrac{4\pi}{a\sqrt{3}}(0,1)
\end{align}
The separation between the kagome sites are given by 
\begin{align}
\mathbf{r}_a-\mathbf{r}_b &=\mathbf{a}_3=\mathbf{r}_a= \tfrac{a}{2}(1,0)\\
\mathbf{r}_a-\mathbf{r}_c &=-\mathbf{a}_2= \mathbf{r}_b=\tfrac{a}{2}(-\tfrac{1}{2},\tfrac{\sqrt{3}}{2}) \\
\mathbf{r}_b-\mathbf{r}_c &=-\mathbf{a}_1=\mathbf{r}_c= \tfrac{a}{2}(-\tfrac{1}{2},-\tfrac{\sqrt{3}}{2})
\end{align}
The locations of the $i$ and $j$ sites are given by
\begin{align}
\mathbf{r}_i=-\mathbf{r}_j &=\tfrac{a}{2}(1,-\tfrac{\sqrt{3}}{4})
\end{align}
and the differences between these sites are given by
\begin{align}
\mathbf{r}_i-\mathbf{r}_j &= \tfrac{a}{2}(2,-\tfrac{\sqrt{3}}{2})\\
\mathbf{r}_i-\mathbf{r}_a &=\tfrac{a}{2}(0,-\tfrac{\sqrt{3}}{4})\\
\mathbf{r}_i-\mathbf{r}_b &= \tfrac{a}{2}(\tfrac{3}{2},-\tfrac{3\sqrt{3}}{4}) \\
\mathbf{r}_i-\mathbf{r}_c &= \tfrac{a}{2}(\tfrac{3}{2},\tfrac{\sqrt{3}}{4})
\end{align}

If we take a simple tight-binding model for the kagome sites with nearest neighbour hopping,
\begin{align}
\mathscr{H} = -t\sum_{\langle i,j\rangle} c^\dag_ic_j + \text{h.c.}
\end{align}
then we get a matrix in sublattice space upon Fourier transforming,
\begin{align}
\label{Htb}
\mathscr{H}(\mathbf{k}) = -\begin{pmatrix}
0&  2t\cos(\mathbf{r}_c\cdot\mathbf{k}) & 2t\cos(\mathbf{r}_b\cdot\mathbf{k})\\ 
 2t\cos(\mathbf{r}_c\cdot\mathbf{k})&  0& 2t\cos(\mathbf{r}_a\cdot\mathbf{k})\\ 
 2t\cos(\mathbf{r}_b\cdot\mathbf{k})&   2t\cos(\mathbf{r}_a\cdot\mathbf{k})& 0
\end{pmatrix}
\end{align}
At each point in momentum space there are three energies i.e. three bands; near the $M$-point, we have the flat band with energy $2t$, as well as the two saddle points with energy 0 in the `conduction band' and $-2t$ in the `valence band'. The Fermi surface is a hexagon with van Hove singularities at the momenta $\mathbf{M}_\gamma$ with $\gamma=1,2,3$ given by 
\begin{align}
\mathbf{M}_1&=\tfrac{2\pi}{a\sqrt{3}}(\tfrac{\sqrt{3}}{2},\tfrac{1}{2}) \\
\mathbf{M}_2&=\tfrac{2\pi}{a\sqrt{3}}(0,-1) \\
\mathbf{M}_3&=\tfrac{2\pi}{a\sqrt{3}}(-\tfrac{\sqrt{3}}{2},\tfrac{1}{2}) 
\end{align}
which give the phases
\begin{align}
&\mathbf{M}_1\cdot \mathbf{r}_a=\tfrac{\pi}{2} && \mathbf{M}_1\cdot \mathbf{r}_b=0  && \mathbf{M}_1\cdot \mathbf{r}_c =-\tfrac{\pi}{2}\\
&\mathbf{M}_2\cdot \mathbf{r}_a=0&& \mathbf{M}_2\cdot \mathbf{r}_b =-\tfrac{\pi}{2} && \mathbf{M}_2\cdot \mathbf{r}_c=\tfrac{\pi}{2}\\
&\mathbf{M}_3\cdot \mathbf{r}_a=-\tfrac{\pi}{2}&& \mathbf{M}_3\cdot \mathbf{r}_b=\tfrac{\pi}{2}  && \mathbf{M}_3\cdot \mathbf{r}_c=0
\end{align}
The Hamiltonian at these momenta is therefore given by
\begin{align}
\mathscr{H}(\mathbf{M}_1) = -2t\begin{pmatrix}
0&  0& 1\\ 
 0&  0& 0\\ 
 1&   0& 0
\end{pmatrix} && \mathscr{H}(\mathbf{M}_2) = -2t\begin{pmatrix}
0&  0& 0\\ 
 0&  0& 1\\ 
 0&   1& 0
\end{pmatrix} && \mathscr{H}(\mathbf{M}_3) = -2t\begin{pmatrix}
0&  1& 0\\ 
 1&  0& 0\\ 
 0&   0& 0
\end{pmatrix}
\end{align}
From these matrices we find the eigenvectors at each patch corresponding to the two saddle points, denoting the conduction/valence band states as $|\gamma,\pm\rangle$ (we shall ignore the flatband). Near $\mathbf{M}_1$ we have
\begin{align}
|1,+\rangle = \begin{pmatrix}
0\\ 
1\\ 
0
\end{pmatrix} , \ \ \ \ \  |1,-\rangle = \tfrac{1}{\sqrt{2}}\begin{pmatrix}
1\\ 
0\\ 
1
\end{pmatrix}
\end{align}
Near $\mathbf{M}_2$ we have
\begin{align}
|2,+\rangle = \begin{pmatrix}
1\\ 
0\\ 
0
\end{pmatrix} , \ \ \ \ \  |2,-\rangle = \tfrac{1}{\sqrt{2}}\begin{pmatrix}
0\\ 
1\\ 
1
\end{pmatrix}
\end{align}
and near $\mathbf{M}_3$ we have
\begin{align}
|3,+\rangle = \begin{pmatrix}
0\\ 
0\\ 
1
\end{pmatrix} , \ \ \ \ \  |3,-\rangle = \tfrac{1}{\sqrt{2}}\begin{pmatrix}
1\\ 
1\\ 
0
\end{pmatrix}
\end{align}
Note that the conduction band saddle point eigenvectors at any given patch have support on only one sublattice, while the valence band saddle point eigenvectors at a given patch have support on two sublattices -- the two types of saddle point are referred to as `$p$-type' (pure) and `$m$-type' (mixed) in the literature, and their sublattice structure can have an important influence on possible instabilities \cite{Kiesel2013, Kiesel2012, ingham2024theory, jiang2024van, wenger2024theory, schwemmer2024sublattice, profe2024kagome, Tazai2022, tazai2023charge}.

The triangular lattice of Sb atoms in V$_3$Sb mean that the monolayer is not simply described by a nearest neighbour kagome tight-binding model, but since the Sb host $p$-orbitals while the V atoms host $d$-orbitals, the two sites have different $M_z$ eigenvalues and so no direct hopping process is possible. We therefore approximate the bilayer as decoupled triangular and kagome lattices. Since the Sb $\Gamma$ pocket behaves trivially under rotations, we therefore also advocate for neglecting the effects of the triangular lattice and focus solely on the kagome lattice of vanadium.

\subsection{Tunneling potential}
\label{cont_deriv}

We now derive the form of the tunneling potential; The derivation follows closely the logic employed in Ref. \cite{Khalaf2019}. Consider the interlayer tunneling between layer $i$ band $a$ and layer $j$ band $b$, in a reference coordinate system
\begin{align}
T^{ij}_{ab} = \int d\mathbf{r}d\mathbf{r'} c^\dag_{i,a}(\mathbf{r}) T^{ij}_{ab}(\mathbf{r},\mathbf{r}')c_{j,b}(\mathbf{r}')
\end{align}
where $c^\dag_{i,a}(\mathbf{r})$ is the full electron creation operator. We now derive the form of the tunneling matrix elements near the vectors $\mathbf{M}_\alpha$ in momentum space. Since the density of states is strongly peaked near the $M$-points, we imagine the electrons are located entirely near patches around $\mathbf{M}_\alpha$. Therefore we only account for tunneling between states at $\mathbf{M}_\alpha$. We also assume the atomic orbitals $\phi^a_{\mathbf{R},\sigma}=\phi^a(\mathbf{r} - \mathbf{R} - \mathbf{r}_\sigma)$ are strongly localised at $\mathbf{r} = \mathbf{R} +\mathbf{r}_\sigma$, where $\mathbf{R}=m_1\mathbf{b}_1+m_2\mathbf{b}_2$ is a lattice site and $\mathbf{r}_\sigma$ is the real space position of the sublattice $\sigma$. Letting $R_{\theta}$ be a $\theta$-rotation matrix, and accounting for a relative displacement $\mathbf{d}_{i}$ between the two layers, we have the fermionic operator in layer $i$,
\begin{align}
c_{i,a}(\mathbf{r})=\sum_{\mathbf{R},\sigma} \phi^a_{\mathbf{R},\sigma}(R_{-\theta}(\mathbf{r}-\mathbf{d}_i))f_{i,\mathbf{R},\sigma,a}=\sum_{\mathbf{R},\sigma} \phi(R_{-\theta}(\mathbf{r}-\mathbf{d}_i)- \mathbf{R} - \mathbf{r}_\sigma)f_{i,\mathbf{R},\sigma,a}
\end{align}
where $f$ creates an electron state corresponding to the centre of an atomic orbital, which we can consider as Wannier functions associated to different bands, ignoring overlaps. We treat the different bands as arising from identical orbitals, ignoring overlap factors. Approximating $\phi(\mathbf{r})\approx\delta(\mathbf{r})$, we obtain
\begin{align}
T^{ij}_{ab} &= \sum_{\mathbf{R},\sigma,\mathbf{R}',\sigma'} T^{ij}_{ab}(\mathbf{d}_i+R_{\theta}(\mathbf{R}+\mathbf{r}_\sigma),\mathbf{d}_j+R_{\theta}(\mathbf{R}'+\mathbf{r}_{\sigma'}))f^\dag_{i,\mathbf{R},\sigma,a}f_{j,\mathbf{R}',\sigma',b}
\end{align}
We assume that $T^{ij}(\mathbf{r},\mathbf{r}') = T^{ij}(|\mathbf{r}-\mathbf{r}'|)$ is translationally and rotationally invariant at the moir\' e scale. We now move to momentum space using the Fourier transforms:
\begin{align}
f_{i,\mathbf{R},\sigma,a} &= \sum_{\mathbf{p}} e^{i\mathbf p \cdot (\mathbf{R}+\mathbf{r}_\sigma)} u_{a,\mathbf{p},\sigma} \psi_{i,\mathbf{p},a} \\
T^{ij}_{ab}(|\mathbf{r}-\mathbf{r}'|) &= \sum_{\mathbf{q}}e^{i\mathbf q \cdot (\mathbf{r}-\mathbf{r}')}T^{ij}_{ab}(\mathbf q)
\end{align}
where $u_{a, \mathbf{p},\sigma}$ is a Bloch function for orbital $a$, which we may think of as effecting a basis transformation from the sublattice basis to band basis,
\begin{align}
T^{ij}_{ab} &= \sum_{\mathbf{R},\sigma,\mathbf{R}',\sigma',\mathbf{p},\mathbf{p}',\mathbf{q}} e^{i\mathbf{q}\cdot (\mathbf{d}_i-\mathbf{d}_j+R_{\theta_i}(\mathbf{R}+\mathbf{r}_\sigma)-R_{\theta_j}(\mathbf{R}'+\mathbf{r}_{\sigma'}))}T^{ij}_{ab}(\mathbf{q})e^{-i\mathbf p \cdot (\mathbf{R}+\mathbf{r}_\sigma)}e^{i\mathbf{p}' \cdot (\mathbf{R}'+\mathbf{r}_{\sigma'})} u_{a,\mathbf{p}',\sigma'} u_{b,\mathbf{p},\sigma} \psi^\dag_{i,\mathbf{p},a}\psi_{j,\mathbf{p}',b}  \\
&=\sum_{\mathbf{R},\sigma,\mathbf{R}',\sigma',\mathbf{p},\mathbf{p}',\mathbf{q}}e^{i\mathbf{R}'\cdot (\mathbf{p}' -R_{-\theta_j}\mathbf{q})}e^{i \mathbf{R} \cdot (-\mathbf p+R_{-\theta_i}\mathbf{q})} e^{i\mathbf{q}\cdot (\mathbf{d}_i-\mathbf{d}_j+R_{\theta_i}\mathbf{r}_\sigma-R_{\theta_j}\mathbf{r}_{\sigma'})}T^{ij}_{ab}(\mathbf{q})e^{-i\mathbf p \cdot \mathbf{r}_\sigma}e^{i\mathbf{p}' \cdot \mathbf{r}_{\sigma'}}u_{a,\mathbf{p}',\sigma'} u_{b,\mathbf{p},\sigma}\psi^\dag_{i,\mathbf{p},a}\psi_{j,\mathbf{p}',b} \nonumber
\end{align}
Since we work with band rather than the sublattice basis, $\sigma$ is summed over. Performing the summation over $\mathbf{R}$ and $\mathbf{R}'$, we find that $-\mathbf p+R_{-\theta_i}\mathbf{q}=\mathbf{G}$ and $\mathbf p'-R_{-\theta_j}\mathbf{q}=\mathbf{G}'$ where $\mathbf{G}$ and $\mathbf{G}'$ are the unrotated reciprocal lattice vectors, which comprise $\mathbf{G}=n_1\mathbf{G}_1 + n_2\mathbf{G}_2$. Clearly we must have $R_{\theta_i}(\mathbf{p}+\mathbf{G})=R_{\theta_j}(\mathbf{p}'-\mathbf{G}')$. The result is
\begin{align}
T^{ij}_{ab} &=\sum_{\sigma,\sigma',\mathbf{p},\mathbf{p}',\mathbf{G},\mathbf{G}'}  e^{iR_{\theta_i}(\mathbf{p}+\mathbf{G})\cdot (\mathbf{d}_i-\mathbf{d}_j)}T^{ij}_{ab}(|\mathbf{p}+\mathbf{G}|)e^{i\mathbf G \cdot \mathbf{r}_\sigma}e^{-i\mathbf{G}' \cdot \mathbf{r}_{\sigma'}}u_{a,\mathbf{p}',\sigma'} u_{b,\mathbf{p},\sigma}\psi^\dag_{i,\mathbf{p},a}\psi_{j,\mathbf{p}',b}
\end{align}
where we used rotational invariance, and bear in mind that $\mathbf{p}'=R_{\theta_i-\theta_j}(\mathbf{p}+\mathbf{G})-\mathbf{G}'$. The operator $\psi^\dag_{i,\sigma,\mathbf{p}}$ creates a Bloch state with quasimomentum $\mathbf{p}$; we expand this state near the momentum $\mathbf{M}_{\alpha}$, in terms of an annihilation operator for the slow varying orbitals at the $\alpha$ patch, $\chi_{i,a,\alpha}(\mathbf{r})$,
\begin{align}
\label{patchorbs}
\psi_{i,\mathbf{p},a} = \int d\mathbf{r}'\  e^{-i(\mathbf{p}-\mathbf{M}_\alpha)\cdot \mathbf{r}'} \chi_{i,\mathbf{k},a,\alpha}(\mathbf{r}')=\int d\mathbf{r}\  e^{-i(\mathbf{p}-\mathbf{M}_\alpha)\cdot R_{-\theta_i}(\mathbf{r}-\mathbf{d}_i)} \chi_{i,a,\alpha}(\mathbf{r})
\end{align}
\begin{figure}[t]
	\includegraphics[height = 0.35\textwidth]{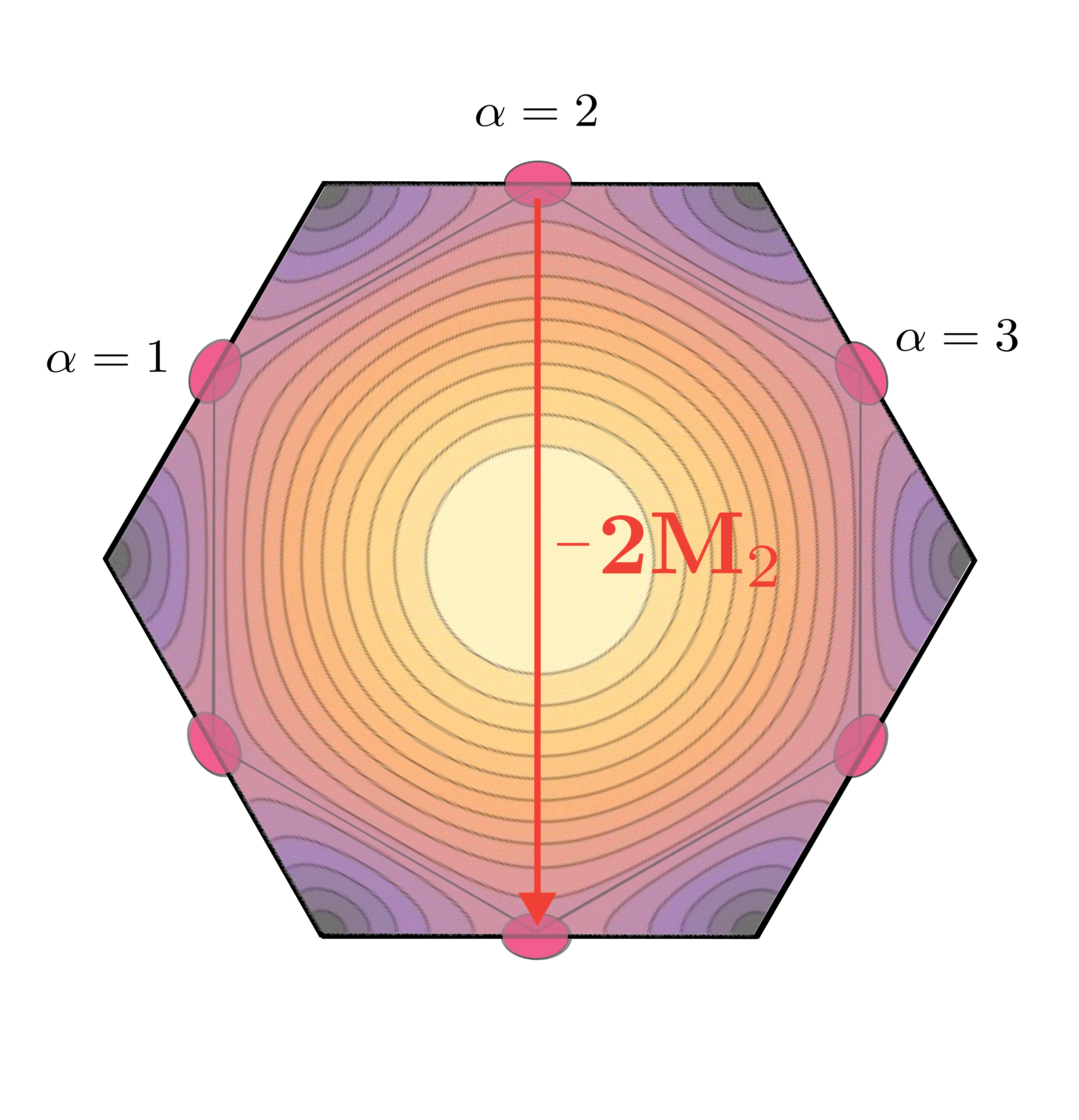}				\vspace{-0.6cm}
	\caption{The minimal set of reciprocal lattice vectors connecting $\mathbf{M}_\alpha$ to equivalent points in the first Brillouin zone are $\mathbf{0}, -\mathbf{2M_\alpha}$, illustrated here for $\alpha=2$.}
	\label{fig:recip}
\end{figure}where we rotated/displaced from the coordinate system in the $i$th layer back to the reference coordinate system. We note that in what follows we can actually work in momentum space, but giving an approximate real space form for the tunneling term will help build intuition. Assuming that $T^{ij}_{ab}(|\mathbf{q}|)$ drops off with large $|\mathbf{q}|$, we keep only the minimal set of $\mathbf{p}+\mathbf{G}$ i.e. we choose the smallest set of momentum points connected to $\mathbf{M}_{\alpha}$ by a reciprocal lattice vector -- which form a set of two points for each patch, as shown in Fig \ref{fig:recip}, namely \{$\mathbf{M}_1$,$\mathbf{M}_1- \mathbf {G}_1$\}, \{$\mathbf{M}_2$, $\mathbf{M}_2- \mathbf {G}_2$\}, and \{$\mathbf{M}_3$, $\mathbf{M}_3- \mathbf {G}_3$\}, where $\mathbf{G}_3=\mathbf{G}_2-\mathbf{G}_1$. From this assumption we also have $\mathbf{G}=\mathbf{G}'$ and $\alpha=\alpha'$ as we must satisfy the equation
\begin{align}
R_{\theta_i}(\mathbf{p}+\mathbf{G})=R_{\theta_j}(\mathbf{p}'+\mathbf{G}')
\end{align}
for $\mathbf{p},\mathbf{p}'$ near the patches. In the leading harmonic apprixmation, the sum over $\mathbf{G}$ is restricted to $\mathbf{G}=0$, and $-2\mathbf{M}_\alpha=- \mathbf{G}_\alpha$, i.e. $- \mathbf{G}_1$, $-\mathbf{G}_2$, or $-\mathbf{G}_3$ depending on the patch. We approximate $T^{ij}_{ab}(|\mathbf{M}_\alpha + \mathbf{G}_n|)_{\sigma\sigma'}\approx w^{\sigma\sigma'}_{ab,n}$ as roughly constant near the patches, where $n$ denote the reciprocal lattice vectors in the $n$th shell. Using \eqref{patchorbs} we can perform the Fourier transform to real space assuming $u_{a,\mathbf{p}',\sigma'} u_{b,\mathbf{p},\sigma}$ is a momentum independent constant; in the case of the kagome lattice, they are given by the sublattice vectors of the previous section. Different bands are orthogonal, so then flavours decouple in addition to the patch decoupling we have already seen.

We arrive at the real space expression
\begin{align}
T^{ij}_{ab}(\mathbf{r}) &=\delta_{\alpha\alpha'}\sum_{\sigma,\sigma',n}   e^{i(R_{\theta_i}-R_{\theta_j})(\mathbf{M}_\alpha+\mathbf{G}_n)\cdot \mathbf{r}_i}\ e^{i\mathbf G_n \cdot (\mathbf{r}_\sigma-\mathbf{r}_{\sigma'})}\ e^{iR_{\theta_i}(\mathbf{M}_\alpha+\mathbf{G}_n)\cdot \mathbf{d}_i-iR_{\theta_j}(\mathbf{M}_\alpha+\mathbf{G}_n)\cdot \mathbf{d}_j} w^{\sigma\sigma'}_{ab,n} u_{a,\sigma'} u_{b,\sigma}
\end{align}
By performing a gauge transformation that absorbs a factor $e^{iR_{\theta}\mathbf{M}_\alpha\cdot \mathbf{d}}$ into the creation operators, this can also be recast as
\begin{align}
\label{continuum_model1}
T^{ij} &=\sum_{\sigma,\sigma',\mathbf{p},\mathbf{p}',n} e^{i(R_{\theta_i}-R_{\theta_j})(\mathbf{M}_\alpha+\mathbf{G}_n)\cdot \mathbf{r}}\ e^{i\mathbf{G}_n\cdot (R_{-\theta_i}\mathbf{d}_i-R_{-\theta_j}\mathbf{d}_j)}\ e^{i\mathbf G_n \cdot (\mathbf{r}_\sigma-\mathbf{r}_{\sigma'})}w^{\sigma\sigma'}_{ab,n} u_{a,\sigma'}u_{b,\sigma}
\end{align}
We now calculate the tunneling matrices explicitly for the case of the kagome lattice. Let us first neglect the band/orbital overlap factor $u_{a,\sigma'} u_{b,\sigma}$. We also assume AA stacking, $\mathbf{d}_i=\mathbf{d}_j=0$. However, we can calculate the results for different stacking arrangements by choosing, for instance, $\mathbf{d}_j=0$ and $\mathbf{d}_i=\mathbf{a}_3, -\mathbf{a}_2$ for AB or AC stacking, or $\mathbf{d}_i=\mathbf{r}_i$ to shift in the way mentioned above, however in our simple model we find a shift in the displacement amounts to a trivial gauge transformation and does not affect the spectrum, so without loss of generality we can set $\mathbf{d}_i=0$.

We then arrive at a simple $M$-point continuum model,
\begin{gather}
T^{ij}_{ab} = \delta_{\alpha\alpha'}\sum_{n} e^{i(R_{\theta_i}-R_{\theta_j})(\mathbf{G}_n+\mathbf{M}_\alpha)\cdot \mathbf{r}}\ T_{ab}(\mathbf{G}_n), \\
T_{ab}(\mathbf{G}_n)_{\sigma,\sigma'} = \sum_{\sigma\sigma'} w^{\sigma\sigma'}_{ab,n} u_{a,\sigma'}u_{b,\sigma} e^{i\mathbf G_n \cdot (\mathbf{r}_\sigma-\mathbf{r}_{\sigma'})}
\end{gather}
Since $\mathbf{M}_\alpha+\mathbf{G} = \mathbf{M}_\alpha,-\mathbf{M}_\alpha$, we find that the moir\' e potential is a one-dimensional modulation $\propto \cos ( (R_{\theta_i}-R_{\theta_j})\mathbf{M}_\alpha\cdot \mathbf{r})$, as explained in the main text.

We now compute the tunneling matrices for the specific case of the kagome lattice in the leading harmonic approximation; we shall also set $w^{\sigma\sigma'}_{ab}=w$, ignoring lattice relaxation effects considering only a single band, and assume $s$-orbitals with trivial sign structure in the tunneling matrix elements. These assumptions are readily extended through a more general form of $w^{\sigma\sigma'}_{ab}$. We begin with the tunneling matrix for $\mathbf{G}=0$. Clearly all the phases are simply zero, so we get
\begin{align}
T(\mathbf{0}) = \begin{pmatrix}
1 & 1 & 1 \\ 
1 & 1 & 1 \\ 
1 & 1 & 1
\end{pmatrix}
\end{align}
Now consider $\mathbf{G}=\mathbf{G}_1$. Using
\begin{align}
\mathbf{G}_1\cdot (\mathbf{r}_a-\mathbf{r}_b) &= \pi\\
\mathbf{G}_1\cdot (\mathbf{r}_a-\mathbf{r}_c) &= 2\pi \\
\mathbf{G}_1\cdot (\mathbf{r}_b-\mathbf{r}_c) &= \pi
\end{align}
we find
\begin{align}
T(\mathbf{G}_1) = \begin{pmatrix}
1 & -1 & 1 \\ 
-1 & 1 & -1 \\ 
1 & -1 & 1
\end{pmatrix}
\end{align}
Now consider $\mathbf{G}=\mathbf{G}_1$. Using
\begin{align}
\mathbf{G}_2\cdot (\mathbf{r}_a-\mathbf{r}_b) &= -\pi\\
\mathbf{G}_2\cdot (\mathbf{r}_a-\mathbf{r}_c) &= \pi \\
\mathbf{G}_2\cdot (\mathbf{r}_b-\mathbf{r}_c) &= 2\pi
\end{align}
we find
\begin{align}
T(\mathbf{G}_2) = \begin{pmatrix}
1 & -1 & -1 \\ 
-1 & 1 & 1 \\ 
-1 & 1 & 1
\end{pmatrix}
\end{align}
Then from 
\begin{align}
\mathbf{G}_3\cdot (\mathbf{r}_a-\mathbf{r}_b) &= -2\pi\\
\mathbf{G}_3\cdot (\mathbf{r}_a-\mathbf{r}_c) &= -\pi\\
\mathbf{G}_3\cdot (\mathbf{r}_b-\mathbf{r}_c) &= \pi
\end{align}
we find
\begin{align}
T(\mathbf{G}_3) = \begin{pmatrix}
1 & 1 & -1 \\ 
1 & 1 & -1 \\ 
-1 & -1 & 1
\end{pmatrix}
\end{align}
Incorporating the orbital overlap factor, for $s$-orbitals at the $p$-type saddle point we get $u_{a,\sigma'}u_{b,\sigma}\propto \delta_{\sigma\sigma'}$ and is non vanishing for only one sublattice. Hence the sublattice sum only changes an overall positive prefactor -- one finds the same in the $m$-type case. A more complex orbital structure can in principle change the tunneling matrix.

As can be seen from Eq. \eqref{continuum_model1}, the moir\' e potential imparts a momentum transfer between the rotated momenta relative to the $M$-point,
\begin{align}
\pm \mathbf{Q}_{\alpha} \equiv R_{\theta_i}(\mathbf{M}_{\alpha}+\mathbf{G}_i)-R_{\theta_j}(\mathbf{M}_\alpha+\mathbf{G}_i)=\pm (R_{\theta_i}\mathbf{M}_{\alpha}-R_{\theta_j}\mathbf{M}_{\alpha})
\end{align}
where $\mathbf{M}_{\alpha}+\mathbf{G}_i$ are simply those momenta equivalent to $\mathbf{M}_{\alpha}$ -- see Fig. \ref{fig:recip} and our previous discussion. Assuming also $\theta_i=-\theta_j=\theta/2$, we find
\begin{align}
\mathbf{Q}_{2}&=R_{\theta_i}\mathbf{M}_{2}-R_{\theta_j}\mathbf{M}_{2}=\tfrac{2\pi}{a\sqrt{3}}\begin{pmatrix}
\sin\theta_i-\sin\theta_j\\
-\cos\theta_i+\cos\theta_j
\end{pmatrix}=\tfrac{2\pi}{a\sqrt{3}}\begin{pmatrix}
2\sin\left(\tfrac{\theta_i-\theta_j}{2}\right)\cos\left(\tfrac{\theta_i+\theta_j}{2}\right)\\
2 \sin \left(\tfrac{\theta _i-\theta_j}{2}\right) \sin \left(\tfrac{\theta
   _i+\theta_j}{2}\right)
\end{pmatrix} =\tfrac{2\pi}{a\sqrt{3}}\begin{pmatrix}
2\sin(\tfrac{\theta}{2})\\
0
\end{pmatrix} 
\end{align}
and $\mathbf{Q}_{1}=R_{-2\pi/3}\mathbf{Q}_{2}$ and $\mathbf{Q}_{3}=R_{2\pi/3}\mathbf{Q}_{2}$ are simply related to $\mathbf{Q}^i_{2}$ by $2\pi/3$ rotations.

Writing the explicit tunneling Hamiltonian for AA stacking in this case,
\begin{align}
T^{ij}(\mathbf{r}) =w \ \delta_{\alpha\alpha'}\cos (\mathbf{Q}_\alpha \cdot \mathbf{r})
\end{align}
Adding a displacement $\mathbf{d}_i=\mathbf{d}$ is easily seen to result simply in $\mathbf{r} \rightarrow \mathbf{r} + \mathbf{d}$, which results in an identical moir\'e dispersion. We therefore conclude that for $M$-point materials, the bandstructure does not depend on the global stacking configuration.

Next we consider the form of the single particle Hamiltonian. In the kagome model tight binding model with nearest neighbour hoppings, we find the dispersions near the $M$-points in dimensionless units of momentum and energy,
\begin{align}
\varepsilon(\mathbf{M}_1+\mathbf{p})&\approx 1 +\tfrac{1}{2}p_x(p_x+\sqrt{3}p_y)+\mathcal{O}(p^2) =1+\varepsilon_{1}(\mathbf{p})\\
\varepsilon(\mathbf{M}_2+\mathbf{p})&\approx 1 +\tfrac{1}{4}(-p_x^2+3p_y^2)+\mathcal{O}(p^2) =1+\varepsilon_{2}(\mathbf{p})\\
\varepsilon(\mathbf{M}_3+\mathbf{p})&\approx 1 +\tfrac{1}{2}p_x(p_x-\sqrt{3}p_y)+\mathcal{O}(p^2)=1+\varepsilon_{3}(\mathbf{p})
\end{align}
It is convenient to write the dispersion in terms of $p_x=p\cos \theta_p$ and $p_x \pm \sqrt{3} p_y= \pm 2p\sin(\theta_p\pm\phi)$
\begin{align}
\varepsilon_{1}(\mathbf{p})&= p^2\cos\theta_p\sin(\theta_p+\phi) \\
\varepsilon_{2}(\mathbf{p})&=  p^2\sin(\theta_p+\phi)\sin(\theta_p-\phi)\\
\varepsilon_{3}(\mathbf{p})&= p^2\cos\theta_p\sin(\theta_p-\phi)
\end{align}
where $\phi=\pi/6$ and $\theta_p=\cos^{-1}(p_x/p)$ is the angle of the momentum $\mathbf{p}$. Monolayer rotation of a momentum $\mathbf{k}$ is then implemented in the dispersion by taking $\theta_k\rightarrow\theta_k+\theta$. 

\begin{figure}[t]
	\includegraphics[height = 0.45\textwidth]{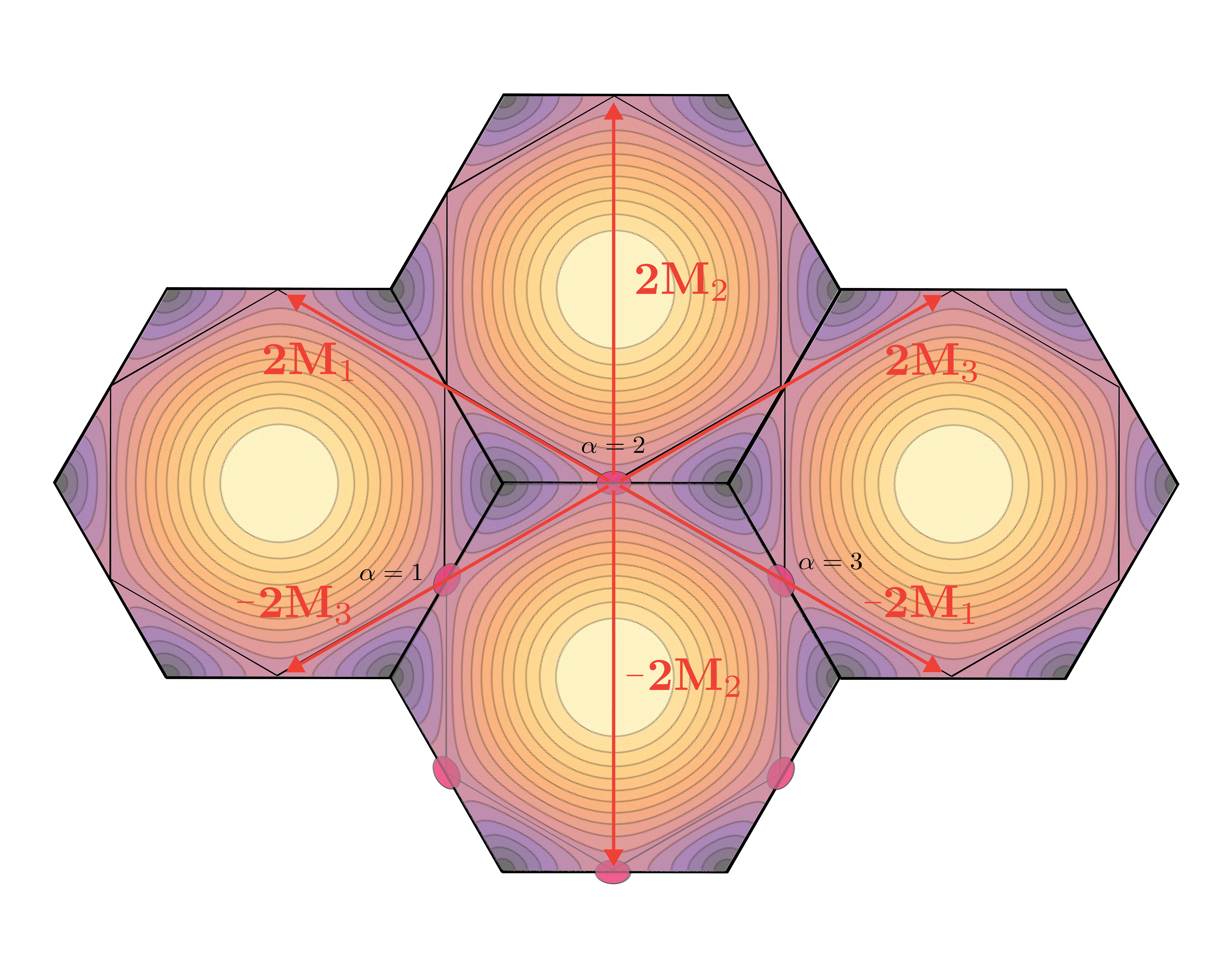}				\vspace{-0.3cm}
	\caption{Subleading momentum transfers involving scattering processes connecting $\mathbf{M}_\alpha$ to equivalent points in momentum space, illustrated for $\alpha=2$. In addition to $\mathbf{0}, -\mathbf{2M_\alpha}$ in the first Brillouin zone, we have $\mathbf{2M_\alpha},\pm \mathbf{2M_{\alpha'}}$ for $\alpha'\neq \alpha$. }
	\label{fig:recip2}
\end{figure}

\subsection{Subleading harmonics in the moir\'e potential}
The leading harmonics in the interlayer potential we derived are one-dimensional at each patch, meaning that one component of momentum is not folded into the moire BZ at each patch. To `regulate' this one may consider the higher harmonics, which should come with significantly suppressed amplitude in the moir\'e potential. The next leading harmonics are simply those that appear in the potential for the other patches, as shown in Fig. \ref{fig:recip2}. As already derived, the leading harmonic at $\alpha=2$ corresponds to
\begin{align}
\mathbf{Q}^1_{2}&=R_{\theta_i}\mathbf{M}_{2}-R_{\theta_j}\mathbf{M}_{2}=\tfrac{2\pi}{a\sqrt{3}}\begin{pmatrix}
\sin\theta_i-\sin\theta_j\\
-\cos\theta_i+\cos\theta_j
\end{pmatrix}=\tfrac{2\pi}{a\sqrt{3}}\begin{pmatrix}
2\sin\left(\tfrac{\theta_i-\theta_j}{2}\right)\cos\left(\tfrac{\theta_i+\theta_j}{2}\right)\\
2 \sin \left(\tfrac{\theta _i-\theta_j}{2}\right) \sin \left(\tfrac{\theta
   _i+\theta_j}{2}\right)
\end{pmatrix} =\tfrac{2\pi}{a\sqrt{3}}\begin{pmatrix}
2\sin(\tfrac{\theta}{2})\\
0
\end{pmatrix}
\end{align}
From Fig. \ref{fig:recip2} one sees that the next leading harmonic is $\pm \mathbf{Q}_2^2$, where
\begin{align}
\mathbf{Q}^2_{2}&=\tfrac{2\pi}{a\sqrt{3}}\left(
\begin{array}{c}
 -2 \sqrt{3} \sin \left(\tfrac{\theta _i-\theta _j}{2}\right) \sin
   \left(\tfrac{\theta _i+\theta _j}{2}\right) \\
 2 \sqrt{3} \sin \left(\tfrac{\theta _i-\theta _j}{2}\right) \cos
   \left(\tfrac{\theta _i+\theta _j}{2}\right) \\
\end{array}
\right)=\tfrac{2\pi}{a\sqrt{3}}\left(
\begin{array}{c}
 0 \\
 2 \sqrt{3}\sin \left(\tfrac{\theta }{2}\right) \\
\end{array}
\right) \equiv \mathbf{Q}'_2
\end{align}
Using the results of the previous section, one finds the associated tunneling matrices are identical for this momentum transfer. Hence, at the level of the first two harmonics, the moir\'e potential at $\alpha=2$ is given by
\begin{align}
    T^{ij}_{ab}(\mathbf{r}) = \delta_{\alpha\alpha'} \left( w_{1ab}\cos(Qx) + w_{2ab}\cos(\sqrt{3} Qy) \right)
\end{align}
 where $Q = 4\pi \sin(\theta/2)/(a\sqrt{3})$. This model is identical to that presented in Eq. (S6.58) of Ref. \cite{cualuguaru2024new}. Given isotropic monolayer states, one expects $w_{1ab} \gg w_{2ab}$, which gives rise to the quasi-one-dimensional physics; the fact that the tunneling potential drops off at larger momenta is therefore crucial to this effect, and the faster the tunneling potential drops off with momentum the more sharply one-dimensional the tunneling profile becomes. As mentioned in a footnote to the main text, $w_{2ab}$ can be relatively enhanced in cases where the monolayer states are more dispersive along the direction of the second harmonic, suppressing the `leading' harmonic in the two-center approximation.
 
 The next leading harmonics have momentum components in both the $k_x$ and $k_y$ directions, and correspond to 
\begin{align}
\mathbf{Q}^3_{2}&=2\mathbf{Q}^1_{2}-\mathbf{Q}^2_{2}\\
\mathbf{Q}^4_{2}&=-\mathbf{Q}^1_{2}\\
\mathbf{Q}^5_{2}&=3\mathbf{Q}^1_{2}\\
\mathbf{Q}^6_{2}&=-\mathbf{Q}^2_{2}\\
\mathbf{Q}^7_{2}&=2\mathbf{Q}^1_{2}+\mathbf{Q}^2_{2}
\end{align}
The wavevectors associated to the moir\'e potentials at the other patches, $\mathbf{Q}^{i}_\alpha$  are related by threefold rotations $\mathbf{Q}^i_{1}=R_{-2\pi/3}\mathbf{Q}^i_{2}$ and $\mathbf{Q}^i_{3}=R_{2\pi/3}\mathbf{Q}^i_{2}$.

\subsection{Moir\'e spin-orbit potential}
\label{supp-rashba}

The symmetries of our system are $C_{3z}$, $C_{2z}$, $C_{2x}$, and time-reversal $\mathscr{T}$. Requiring that the Hamiltonian near the three patches are related by threefold rotation automatically satisfies $C_{3z}$. The action of the remaining three symmetries on the spin-orbit potential $\mathscr{S}(\mathbf{r})$ are
\begin{align}
    C_{2z}:& \ \ \ \ \ \ \ \ -i s_z \mathscr{S}(-\mathbf{r}) is_z\\
    C_{2x}:&  \ \ \ \ \ \ \ \  -i \ell_x s_x \mathscr{S}(\mathbf{r}') i \ell_xs_x \\
    \mathscr{T}:& \ \ \ \ \ \ \ \ \ \ \ \, i s_y \mathscr{S}^*(\mathbf{r}) i s_y 
\end{align}
where $\mathbf{r}'=(x,-y)$ and $ \mathscr{S}^*(\mathbf{r})$ indicates complex conjugation. The requirement of $C_{2z}$ symmetry is that any odd function of the coordinates must be accompanied by a matrix which anticommutes with $s_z$, and even functions with a matrix that commutes. As a simple corollary, we conclude that any spin-rotation invariant interlayer moiré coupling like the one in Eq.~\ref{tunnel2d} has to be inversion symmetric, as observed in Ref. \cite{cualuguaru2024new}. This is different for the terms that are non-trivial in spin space. Upon noting that time-reversal $\mathscr{T}$ requires that only $\ell_y$ may appear for those types of terms and further imposing $C_{2z}$ and $C_{2x}$, we find that the most general symmetry-allowed moir\'e spin-orbit potential for $\alpha=2$ reads as
\begin{align}
    \mathscr{S}(\mathbf{r}) = v_1 \sin(Qx) \ell_y s_y + u_1 \cos(Q x) \ell_y s_z + v_2 \sin(\sqrt{3}Qy) \ell_y s_x + + u_2 \cos(\sqrt{3}Qy) \ell_y s_z 
\end{align}
where we have included the first two leading harmonics. In the main text, we neglected the couplings $v_2$ and $u_2$, on the same basis that we neglected $w_2$, which is also associated with the subleading harmonic. 
\newpage

\section{Bandstructure details}

In this section we provide further conceptual discussion of the continuum model bandstructure, and present supplementary plots for a range of parameters beyond those shown in the main text.

\subsection{Action of symmetries on the moir\'e bands}
\label{syms}

To better understand the structure of the moir\'e bands and in particular their spin texture, we discuss the realisation of the symmetries in the moir\'e theory. Some of those symmetries involve a non-zero momentum shift which results from the fact that the $M$ points of the individual layers are located at finite momenta $R_{\ell \theta/2}(\mathbf{M}_\alpha)$ with respect to the untwisted coordinate system (here $\ell=\pm$ indexes the two layers). Using the same conventions for the momentum coordinates $\boldsymbol{k}$ of the low-energy continuum model as in other parts of the paper, where one vHS is located at $\boldsymbol{k}=0$ and the other at $\boldsymbol{k}=(Q,0)^T$ [see, e.g., Fig.~\ref{fig:supp-1Q-plots}(a)], the symmetries are represented on the states $\ket{\boldsymbol{k}}$ as follows: 
\begin{enumerate}
    \item Time-reversal symmetry is realised as usual via
\begin{align}
\mathscr{T}|\mathbf{q}, \ell, s\rangle = is_y|-\mathbf{q}, \ell, s\rangle
\end{align}
\item On the other hand $C_{2x}$ is realised rather unusually, in a non-local fashion in momentum space:
\begin{align}
C_{2x}|\mathbf{q}, \ell, s\rangle = s_x\ell_x|\mathbf{q}', \ell, s\rangle
\end{align}
where $\mathbf{q}' = (q_x+Q,-q_y)$,
\item 
 Similarly $C_{2y}$ is realised as,
\begin{align}
C_{2y}|\mathbf{q}, \ell, s\rangle = s_y\ell_x|-\mathbf{q}', \ell, s\rangle.
\end{align}
\end{enumerate}
Examining Fig.~\ref{fig:cont_disp} in the main text, and the band structure plots to follow in this supplementary section, e.g. Fig. \ref{fig:supp-splitting}, shows that shifting the band structure by $Q$ in $k_x$, reflecting in $k_y$, and flipping the spin indeed is a symmetry.

\subsection{Analytic perturbation theory and $\mathbf{k}\cdot \mathbf{p}$ model}
\label{supp-heff}

In Section \ref{svl} we stated the effective Hamiltonian for weak superlattice and weak moir\'e-Rashba potentials; we include the derivation here. We begin with full Hamiltonian, written as a degenerate perturbation theory in the superlattice potential terms, for the leading harmonic and leading set of k-points,
\begin{align}
\label{H_deg}
    {\cal H}&=\begin{pmatrix}
    \varepsilon^{(22)}(\mathbf p-\mathbf Q) && ( {W}_{\mathbf Q})^{(21)} && 0 \\
    ( {W}_{-\mathbf Q})^{(12)} && \varepsilon^{(11)}(\mathbf p) && ( {W}_{\mathbf Q})^{(12)}\\
    0 && ( {W}_{-\mathbf Q})^{(21)} && \varepsilon^{(22)}(\mathbf p + \mathbf Q)
    \end{pmatrix}
\end{align}
superscripts indicate the layer quantum numbers, and the tunnel matrices are
\begin{align}
     {W}_{\mathbf Q}&= w_1 s_0 \ell_x + i v_1 s_y \ell_y + u_1 s_z \ell_y\\
     {W}_{-\mathbf Q}&= w_1 s_0 \ell_x - i v_1 s_y \ell_y + u_1 s_z \ell_y
\end{align}
and we have denoted the rotated monolayer dispersions as
\begin{align}
\varepsilon^{(11)}(\mathbf p) &= \varepsilon_{2}(R_{\theta/2}\mathbf p)\\
\varepsilon^{(22)}(\mathbf p) &= \varepsilon_{2}(R_{-\theta/2}\mathbf p)
\end{align}
One can explicitly see the action of $C_{2x}$ from \eqref{H_deg}, i.e. shifting $\mathbf p\to \mathbf p+\mathbf Q$ and swapping layer indices, which holds upon extended the Hilbert space to include to all $\mathbf{p}+n \mathbf Q$. 

To arrive at the low-energy $\mathbf k\cdot \mathbf p$ model, due to the non-local action of $C_{2x}$ in $k$-space, it will prove necessary to perform a perturbative expansion about $\mathbf p=\mathbf k$ and $\mathbf p=\mathbf Q+\mathbf k$ with $\mathbf k\approx \mathbf 0$. These expansions are
\begin{align}
{\cal H}'(\mathbf k)&=\varepsilon^{(11)}(\mathbf k) - ( {W}_{\mathbf Q})^{(12)}\frac{1}{\varepsilon^{(22)}(\mathbf k + \mathbf Q)-\varepsilon^{(11)}(\mathbf k)}( {W}_{-\mathbf Q})^{(21)} - ( {W}_{-\mathbf Q})^{(12)}\frac{1}{\varepsilon^{(22)}(\mathbf k - \mathbf Q)-\varepsilon^{(11)}(\mathbf k)}( {W}_{\mathbf Q})^{(21)}\\
{\cal H}'(\mathbf Q + \mathbf k)&=\varepsilon^{(22)}(\mathbf k) - ( {W}_{\mathbf Q})^{(21)}\frac{1}{\varepsilon^{(11)}(\mathbf k + \mathbf Q)-\varepsilon^{(22)}(\mathbf k)}( {W}_{-\mathbf Q})^{(12)} - ( {W}_{-\mathbf Q})^{(21)}\frac{1}{\varepsilon^{(11)}(\mathbf k - \mathbf Q)-\varepsilon^{(22)}(\mathbf k)}( {W}_{\mathbf Q})^{(12)}
\end{align}
Introducing $\eta=\pm$,
\begin{align}
    \sum_{\eta=\pm}( {W}_{\eta \mathbf Q})^{(12)}\frac{1}{\varepsilon^{(22)}(\mathbf k + \eta \mathbf Q)-\varepsilon^{(11)}(\mathbf k)}( {W}_{-\eta \mathbf Q})^{(21)} & = \sum_{\eta=\pm}\frac{1}{2}\left[\gamma_0 s_0 + \eta(\gamma_1 s_x+\gamma_2 s_y)\right] \left[\alpha_0 + \eta(\alpha_1 k_x + \alpha_2 k_y)\right]\\
    &=\alpha_0\gamma_0 s_0 + (\alpha_1 k_x + \alpha_2 k_y)(\gamma_1 s_x + \gamma_2 s_y)\\
    \sum_{\eta=\pm}( {W}_{\eta \mathbf Q})^{(21)}\frac{1}{\varepsilon^{(11)}(\mathbf k + \eta \mathbf Q)-\varepsilon^{(22)}(\mathbf k)}( {W}_{-\eta \mathbf Q})^{(12)} & =\sum_{\eta=\pm} \frac{1}{2}\left[\gamma_0 s_0 + \eta(\gamma_1 s_x-\gamma_2 s_y)\right] \left[\alpha_0 + \eta(\alpha_1 k_x - \alpha_2 k_y)\right]\\
    &=\alpha_0\gamma_0 s_0 + (\alpha_1 k_x - \alpha_2 k_y)(\gamma_1 s_x - \gamma_2 s_y)
\end{align}
Lastly note that $\varepsilon^{(11)}(\mathbf k)=\varepsilon^{(22)}(\mathbf k) + O(k^2) \equiv \varepsilon(\mathbf k)$. We arrive at 
\begin{gather}
{\cal H}_{k\cdot p}=\varepsilon(\mathbf k)  + \alpha_0\gamma_0 s_0 + (\alpha_1 k_x +  \alpha_2 k_y{\ell}_z)(\gamma_1 s_x +  \gamma_2 s_y{\ell}_z) - {\mathbf B} \cdot {\mathbf s}
\label{Hkp}
\end{gather}
Note $C_{2x}$ has the action: ${\ell}_z\to-{\ell}_z$, $s_x\to s_x$, $s_y\to-s_y$ and $(k_x,k_y)\to (k_x,-k_y)$, which we explicitly see to leave $\mathcal{H}_{k\cdot p}$ invariant. Here we have included an in-plane Zeeman field $\mathbf B$ in units $g\mu_B=1$, and we have defined $\gamma_0=(w_1^2+v_1^2+u_1^2)$, $\gamma_2 =2 w_1 v_1$ and  $\gamma_1 =2 v_1 u_1$. Since we expect $w_1>v_1\approx u_1$  the hierarchy is $\gamma_0>\gamma_2>\gamma_1$. For completeness sake we state the expressions for $\alpha$,
\begin{align}
    \alpha_0=-\frac{8}{t Q^2 (2 \cos \theta-1)}, \quad
    \alpha_1 = \frac{2 \alpha_0}{Q}, \quad
    \alpha_2 = \frac{4 \sin\theta \alpha_0}{Q (1-2 \cos\theta)},
\end{align}
recalling that $Q=|\bm Q_2|$ and that $t Q^2$ has dimension of energy. 

To connect the action of $C_{2x}$ in the $\mathbf{k}\cdot\mathbf{p}$ Hamiltonian to the non local realisation described in Sec. \ref{syms}, note that flipping the two layers $\ell\rightarrow -\ell$ also results in a shift of momentum by $\mathbf{Q}$, since the basis of monolayer states are taken at different momenta to those in the top layer.

\subsection{Band structure calculations}

We calculate the continuum model band structure using standard means; to calculate the dispersion for the valley $\alpha$ in the single-harmonic case, we take a basis of states in the top layer $\{e^{i(\mathbf{k}+2n\mathbf{Q}_\alpha) \cdot \mathbf{r}}\}$, and in the bottom layer $\{e^{i(\mathbf{k}+(2n+1)\mathbf{Q}_\alpha )\cdot \mathbf{r}}\}$; we refer to the states indexed by $n$ as k-points. Incorporating the second harmonic expands the basis to $\{e^{i(\mathbf{k}+n\mathbf{Q}_\alpha+m\mathbf{Q}'_\alpha) \cdot \mathbf{r}}\}$ in the top layer, and $\{e^{i(\mathbf{k}+(2n+1)\mathbf{Q}_\alpha+2m\mathbf{Q}'_\alpha )\cdot \mathbf{r}}, e^{i(\mathbf{k}+2n\mathbf{Q}_\alpha+(2m+1)\mathbf{Q}'_\alpha )\cdot \mathbf{r}}\}$ in the bottom layer, where near the patch $\alpha=2$ we have $\mathbf{Q}_2=(Q,0)$ and $\mathbf{Q}'_2 = (0,\sqrt{3}Q)$, and $Q=4\pi \sin(\theta/2)/a\sqrt{3}$. We choose units of momentum so that $Q=1$ and energy so that $tQ^2=1$, after which the Hamiltonian depends on the dimensionless parameters $(w_i/tQ^2, v_i/tQ^2, u_i/tQ^2)$. Note that $2\mathbf{Q}_\alpha$, $2\mathbf{Q}'_\alpha$, $\mathbf{Q}_\alpha+\mathbf{Q}'_\alpha$ are the moir\'e reciprocal lattice vectors, forming a diamond mini Brillouin zone boundary at each valley. When computing the bandstructure for the leading harmonic only, we will often plot the dispersion in the extended zone scheme along $k_y$, i.e. not include the effect of band-folding with respect to $2\mathbf{Q}'_\alpha$ and $\mathbf{Q}_\alpha+\mathbf{Q}'_\alpha$. 

\subsubsection{Tunneling strength and twist-angle dependence}

We now plot the two-dimensional dispersion and bandstructure cuts for a range of parameters not shown in the main text. To begin with, we highlight the features of the bandstructure along the $\Gamma$YX$\Gamma$X line, where these high-symmetry points are those of the moir\'e Brillouin zone.

\begin{figure*}[t!]
	\includegraphics[width = \textwidth]{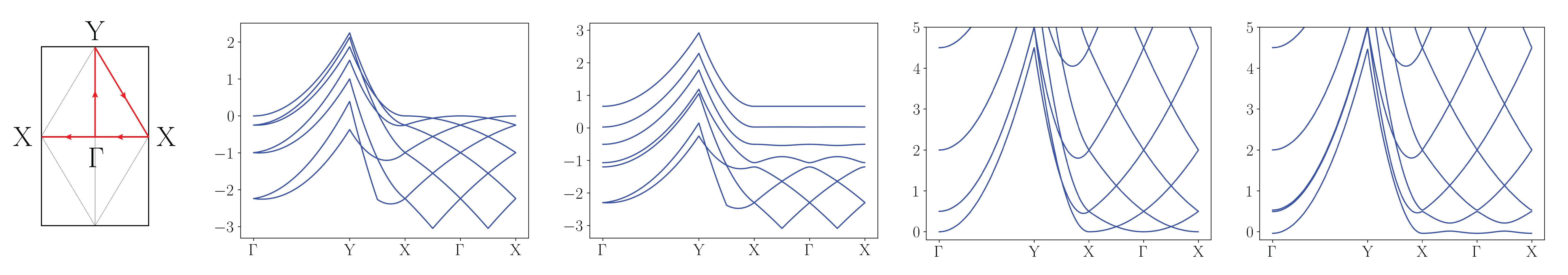}		
    \vspace{-0.7cm}
	\caption{\textbf{Cut bandstructure in the single harmonic approximation.} Cut bandstructure for the single harmonic superlattice bands for the saddle point dispersion (far and middle left) and an ordinary anistropic quadratic band (far and middle right), with no tunneling $w_1=0$ (far left and middle right) and $w_1=0.3,0.1$ (middle left and far right). The plotted bands are projected into a single valley near $\mathbf{Q}_2$, as elsewhere.}
    \vspace{-0.35cm}
	\label{fig:supp-cut-eg-plots}
\end{figure*}

In Fig. \ref{fig:supp-cut-eg-plots}, we show the bandstructure of the single harmonic model with no spin-orbit coupling -- i.e. including only the effects of $w_1$ -- for $\theta=4.25^\circ$ and $w_1=0$ (far left) and $w_1=0.5$ (middle left). We take twelve k-points in each layer, $n=-6,...,6$. One observes explictly that the dispersion is not changed along $k_y$, and at finite $w_1$ the crossings along the $k_x$ axis open up into superlattice gaps. To compare the with the case of a simple quadratic dispersion, we reverse the sign of dispersion along the hole-like saddle direction -- in this case e.g. the dispersion is $\varepsilon_2(\mathbf{k}) = \tfrac{1}{2}k_x^2+\tfrac{3}{2}k_y^2$ near $\mathbf{M}_2$ -- and plot the bands which result in this case for comparison, with $w_1=0$ (middle right) and $w_1=0.1$ (far right).

\begin{figure*}[t!]
	\includegraphics[width = \textwidth]{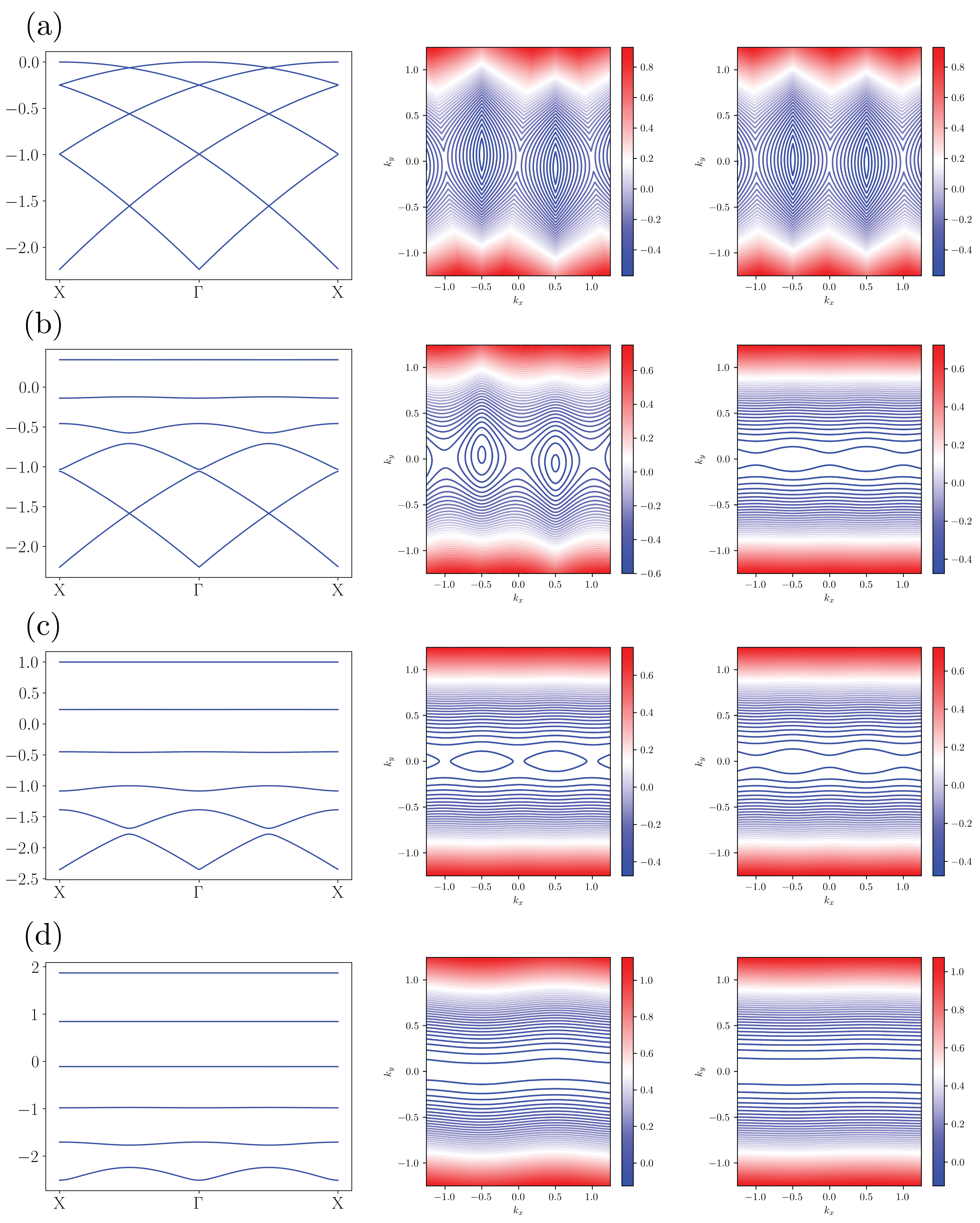}		
	\caption{\textbf{Continuum model bandstructure with zero spin-orbit coupling in the single harmonic approximation.} Single harmonic bandstructure, for (a) $w_1=0$, $w_1=0.3$, $w_1=0.7$, $w_1=1.2$. Left column: bandstructure cut along the X$\Gamma$X line. Middle column: contour plot of the third eigenvalue for $\theta=4.25^\circ$, Right column: contour plot of the third eigenvalue for $\theta=1.05^\circ$.}
	\label{fig:supp-1Q-plots}
\end{figure*}

To illustrate the two-dimensional dispersion, in Fig. \ref{fig:supp-1Q-plots} we show a contour plot of the eigenvalues for several values of the tunneling strength and twist angle. The non locality of the $C_{2x}$ symmetry derived in Sec. \ref{syms} is perhaps most noticeable in panel (b) for the larger twist angle (middle column), in which the pockets at $k_x>0$ and $k_x<0$ are evidently displaced relative to each other in $k_y$.

\subsubsection{Moir\'e-Rashba spin-orbit dependence}

\begin{figure*}[t!]
	\includegraphics[width = \textwidth]{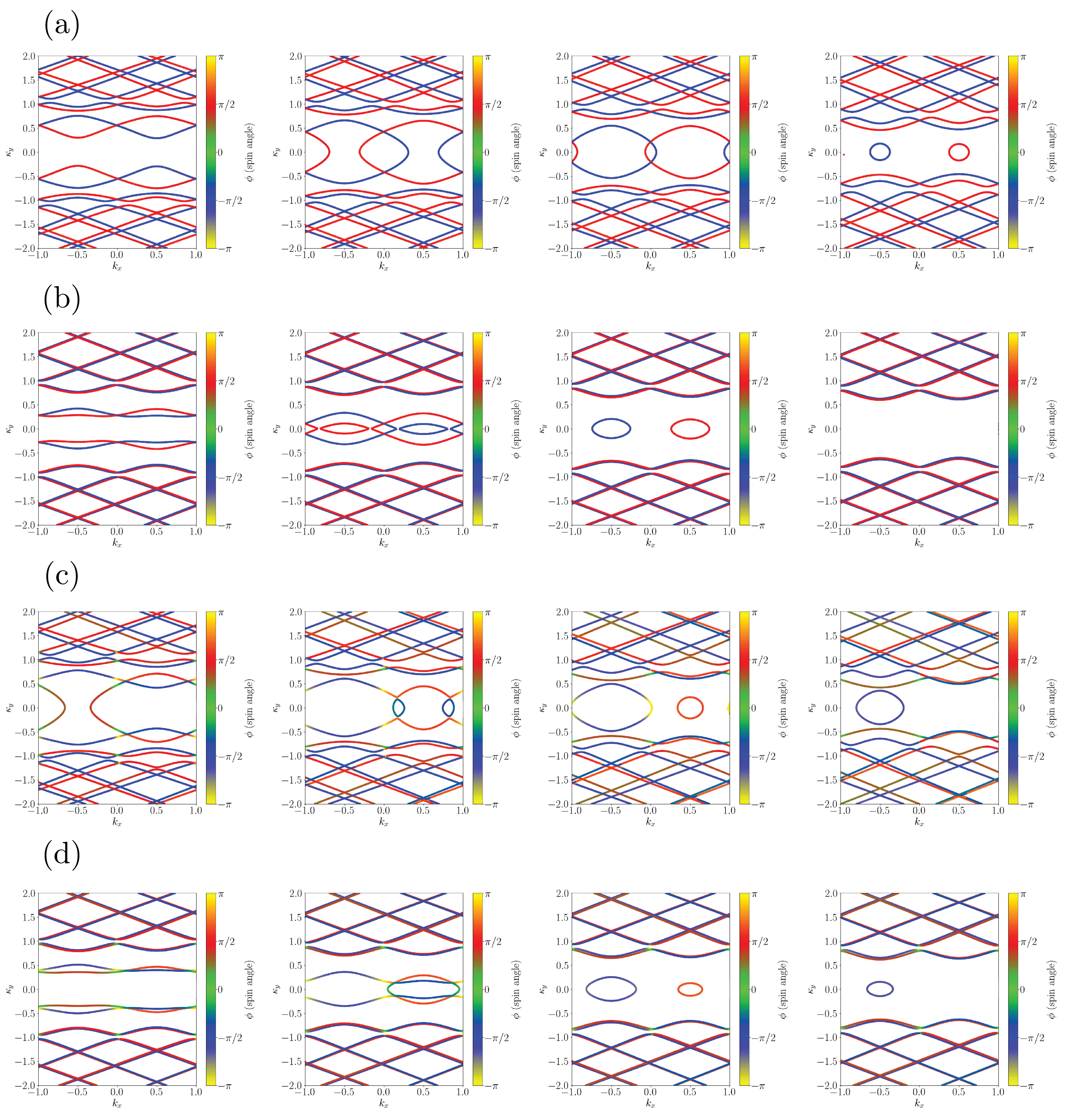}		
	\caption{\textbf{Continuum model bandstructure with various spin-orbit coupling strengths and Zeeman fields.} Energy contours for (a) strong spin-orbit: $E_F=\{-0.3,-0.4,-0.5,-0.7\}$ from left to right, with $w_1=0.5$, and $v_1=u_1=0.25 w_1$, and twist angle $\theta=1.05^\circ$. (b) weak spin-orbit: $E_F=\{-0.45,-0.5,-0.55,-0.6\}$, with $w_1=0.5$, and $v_1=u_1=$ $0.05 w_1$; (c)  strong spin-orbit and strong field: $E_F=\{-0.3,-0.5,-0.6,-0.7\}$, with $w_1=0.5$, and $v_1=u_1=0.25 w_1$ and $B_y=0.125$; (d) weak spin-orbit and weak field: $E_F=\{-0.4,-0.5,-0.55,-0.58\}$, with $w_1=0.5$, and $v_1=u_1=0.05 w_1$ and $B_y=0.025$. Plots are shown for the vicinity of $\mathbf{M}_2$.}
	\label{fig:cont_disp_supp}
\end{figure*}

In Fig. \ref{fig:cont_disp_supp}, we show a range of additional spin-orbit strengths and Zeeman fields. Since the moir\'e-Rashba potential acts as a tunneling term with momentum transfer $\pm\bm Q_2$, bands folded from more distant Brillouin zones experience a relatively weaker spin-orbit than bands originating in the first Brillouin zone. This becomes evident in Fig. \ref{fig:cont_disp_supp}, whereby the bands located at small $|k_y|$ experience a significant spin-splitting, whereas the criss-cross bands, which have been folded several times, show small splitting. 

Including a magnetic field induces spin and momentum polarisation. This effect is most easily seen in the regime of closed moir\'e Fermi surfaces, as opposed to the open, quasi-1D Fermi surfaces. In Fig. \ref{fig:cont_disp_supp} we present a spin and momentum polarisation at various Fermi energies. 

For appropriately chosen parameters, one can create a situation in which one spin species is confined into one-dimensional channels, while the other is unconfined, producing coexisting one- and two-dimensional spin-polarised quasiparticles.  From Fig. \ref{fig:cont_disp_supp}d, we see a transition of the spin-polarised Fermi surfaces as a function of chemical potential:  first, there is a quasi-1D spin-polarised band, then there are both quasi-1D and 2D bands, with opposite spin polarization, finally both spin species are 2D.

To demonstrate spin and momentum polarisation, we direct $\mathbf B$ along $\hat{y}$ instead of $\hat{x}$, since this will correspond to the largest component of SOC, i.e. $\gamma_2>\gamma_1$ in \eqref{main-Hkp}, and hence this direction of magnetic field is most effective at generating a spin and momentum polarisation of the Fermi surfaces. 

As stated in the main text, the direction of the confined spin projection varies from patch to patch by threefold rotation; in real space, this manifests as three sets of parallel quasi-one-dimensional channels, relatively rotated by $2\pi/3$. Twist angle can be tuned to interpolate between weak and strong quasi-one-dimensional confinement, allowing experiments to explore the crossover between these two regimes. In forthcoming work, we visualise the coexisting one- and two-dimensional density of states in real space through simulations of the local density of states, c.f. \cite{kreisel2021quasi, Calugaru2022, Hong2022, sobral2023machine, nag2024pomeranchuk, holbrook2024real}.

\begin{figure*}[t!]
    \centering
    \includegraphics[width=\textwidth]{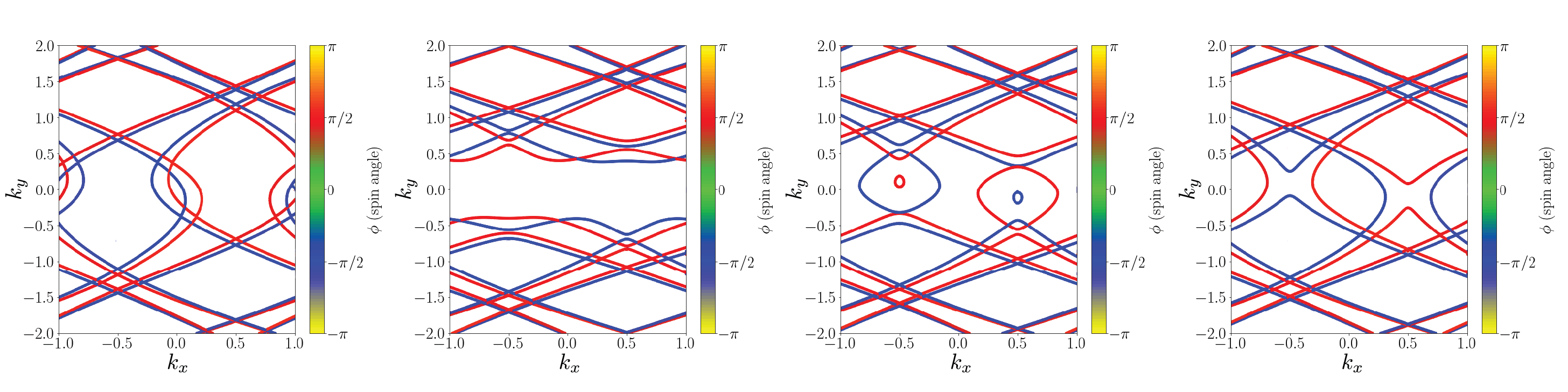}
    \vspace{-0.7cm}
    \caption{\textbf{Spin-splitting at larger twist angles.} Continuum model bandstructure in the vicinity of $\mathbf{M}_2$, with $E_F=\{-0.9,-0.3,-0.5,-0.7\}$ from left to right. Tunneling and spin-orbit matrix elements are taken to be $w_1=v_1=u_1=0.25 w_1$, with twist angle $\theta=10.5^\circ$.}
    \label{fig:supp-splitting}
    \vspace{-0.3cm}
\end{figure*}

In the main text plots, the spin-splitting was negligible along $k_x=0$.  Consulting ${\cal H}_{k.p}$ one sees that there should be spin-splitting at $k_x=0$, and that it is proportional to $\alpha_2$. We noted that the splitting is {\it smaller} than for $k_x\neq0$ in the small twist-angle-limit since $\alpha_2/\alpha_1 \propto \sin\theta$. For completeness, Fig. \ref{fig:supp-splitting} plots the band structure of the continuum model at large twist-angle, $\theta=10.5^\circ$. Indeed, a clear splitting along $k_x=0$ is seen. 

We reiterate that the peculiar non-local in $k$-space action of $C_{2x}$ is quite easily seen in Fig. \ref{fig:supp-splitting}; shifting the band structure by $Q$ in $k_x$, reflecting in $k_y$, and flipping the spin is readily seen to be a symmetry of the bands.

\subsubsection{Inclusion of the subleading harmonics and dense moir\'e bands}

\begin{figure*}[t!]
    \centering
    \includegraphics[width=\textwidth]{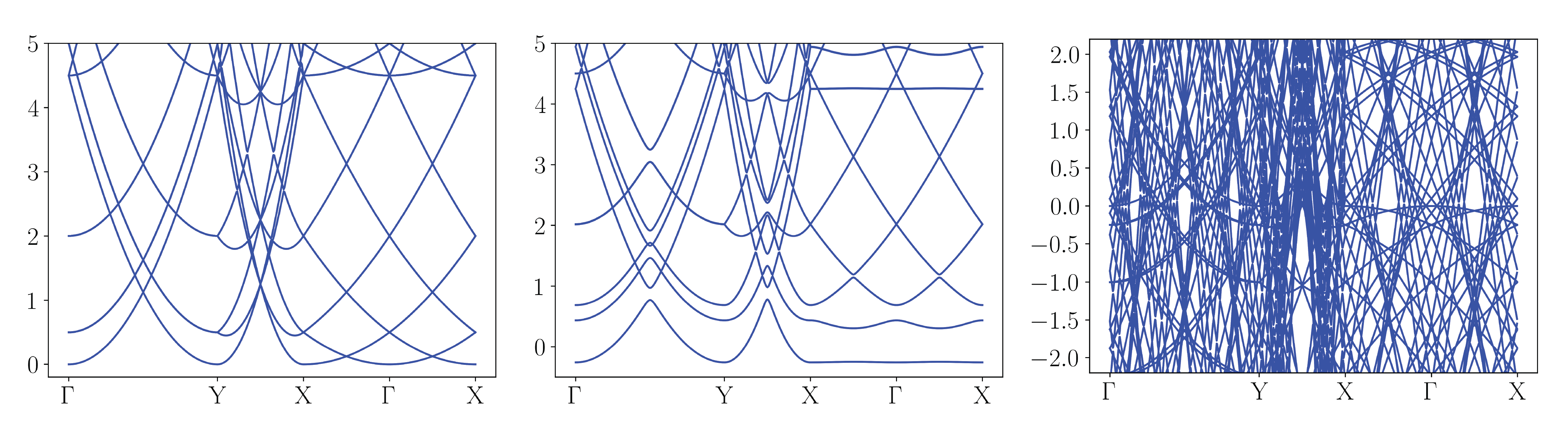}
    \vspace{-0.7cm}
    \caption{\textbf{Influence of the second harmonic.} Cut bandstructure for the monotonic quadratic monolayer dispersion $\varepsilon_2 = \tfrac{1}{2}t(k_x^2+3k_y^2)$, with the inclusion of the next-leading harmonic in the moir\'e potential $w_2$, producing a two-dimensional band reconstruction, for (a) zero tunneling strength, and (b) $w_1=0.3$ and $w_2=0.1$. (c) Bandstructure for the vHS monolayer dispersion $\varepsilon_2 = \tfrac{1}{2}t(k_x^2-3k_y^2)$, producing a dense set of overlapping moir\'e bands around zero energy. }
    \label{fig:second_harmonic}
    \vspace{-0.3cm}
\end{figure*}

To illustrate the bandstructure in the presence of the subleading harmonic, in Fig. \ref{fig:second_harmonic}a and b, we plot the dispersion along the $\Gamma$YX$\Gamma$X line for the case of the monotonic quadratic dispersion used in the previous section, where the pathology of dense moir\'e bands from the vHS is avoided.

In Sec. \ref{twist_vhs} of the main text, we described the band-folding of a large number of degenerate monolayer states near a vHS when harmonics along both directions are included. In this section we show bandstructure plots which illustrate this phenomenon. In Fig. \ref{fig:second_harmonic}c, we show the bandstructure along the $\Gamma$YX$\Gamma$X line restricted to an energy window centered around zero, for zero superlattice potential or spin-orbit coupling -- i.e. we simply band-fold the dispersion, taking XYZ k-points. We note in particular the large density of band crossings along the XY line. With increased superlattice potential, one can discern the opening of superlattice gaps. In forthcoming work, we analyse the structure of interaction form factors within this dense set of bands making efficient use of the adjoint representation, as in \cite{li2020artificial, li2021higher, scammell2022intrinsic, ingham2023quadratic, guerci2024topological}.

\end{document}